\documentclass[runningheads]{llncs}
\usepackage{subcaption}
\usepackage{stmaryrd}
\usepackage{xcolor}
\usepackage{tikz}
\usetikzlibrary{tikzmark,shapes.misc}
\usepackage{color}
\usepackage{wrapfig}
\usepackage{graphicx,wrapfig,lipsum}
\usepackage{float} 

\newcommand{\xmark}{\ding{55}}%
\usepackage{makecell} 

\usepackage{graphicx}
\usepackage{stmaryrd}
\usepackage{amsmath}
\usepackage{amsfonts}
\usepackage{relsize}
\usepackage{amssymb}
\usepackage{mdframed}
\usepackage{comment}
\usepackage{adjustbox}
\usepackage{listings}
\usepackage[most]{tcolorbox}
\usepackage{inconsolata}
\definecolor{mGray}{rgb}{0.5,0.5,0.5}
\usepackage{prooftree}
\usepackage{enumitem}
\usepackage{pifont}
\usepackage[numbers]{natbib}

\newtcblisting[auto counter]{codelisting}[2][]{sharp corners, 
,colframe=gray, listing only, listing options={basicstyle=\tiny\ttfamily,language=java}, 
}

\usepackage{dutchcal}

\newcommand\notationlink[2]{\hyperlink{#1}{\normalcolor #2}}
\newcommand\notationtarget[2]{{\hypertarget{#1}{}}#2}
\newcommand*{\notation}[2]{%
  \expandafter\newcommand\csname #1\endcsname{\notationlink{#1}{#2}}%
  \expandafter\newcommand\csname #1Def\endcsname{\notationtarget{#1}{}}}

\newcommand{\code}[1]{{\tt{\ensuremath{\m{#1}}}}}
\newcommand{\m}{\mathit}  

\newcommand*{\heifer}{\textsc{H}eifer}

\newcommand*{\LoC}{{\scriptsize LoC}}
\newcommand*{\LoS}{{\scriptsize LoS}}
\notation{emp}{\code{emp}}
\notation{req}{\coden{{\bf req}}}
\notation{ens}{\coden{{\bf ens}}}
\newcommand{\vbar}{\code{\bf~|~}}

\newcommand{\hide}[1]{}

\newcommand{\toelim}[1]{}

\newcommand{\seq}{;}%
\newcommand{\spaceT}{~}%

\newcommand{\false}{\m{false}}

\newcommand{\neffectcall}[1]{#1}

\newcommand{\s}[1]{\{ #1 \}}

\newcommand{\nownsay}[1]{}
\newcommand{\wnnay}[1]{}

\notation{HSSL}{\emph{HSSL}}

\notation{pointsto}{\mapsto}

\newcommand{\pttoA}[3]{\ensuremath{{{#1}}\,{\pointsto}\,{{#2}}{{#3}}}}
\newcommand{\heapto}[2]{\pttoA{#1}{}{#2}}

\newcommand{\incrRule}[3]{
 \begin{array}{c}
 \frac{
 \begin{array}{c}
 #1
 \end{array}
 }{
 \begin{array}{c}
 #2
 \end{array}
 }
~\defrulen{#3}
\end{array}
}

\newcommand{\incrHRule}[3]{
\[
 \incrRule{#1}{#2}{#3}
\]
}

\notation{entS}{\,\coden{{\vdash}\,}}
\notation{equivTo}{\coden{\,{<}\!{=}\!{=}\!{>}}}
\notation{normsTo}{\coden{{=}\!{=}\!{>}}}

\newcommand{\HTriple}[3]{\{~#1~\}~#2~\{~#3~\}}

\newcommand{\ml}[1]{\ensuremath{\m{#1}}}

\newcommand{\shortarrow}{\clipbox*{{.38\width} 0pt {\width} {\height}} \textrightarrow}

\newcommand{\nil}{\btt{null}}
\notation{sep}{\,\code{*}\,}
\newcommand{\coden}[1]{\ensuremath{#1}}
\newcommand*{\lam}{\m{fun}}

\newcommand{\codelamsp}[3]{\ensuremath{\lam\,#1\,{#2}\,\shortarrow\,#3}}
\notation{logicallambda}{\lambda}
\newcommand{\codelamspB}[3]{\ensuremath{\logicallambda\,#1\,\shortarrow\,{#2}}}

\newcommand{\btt}[1]{{\ensuremath{\tt #1}}}

\newcommand{\myit}[1]{\textit{#1}}

\notation{xpure}{\myit{xpure}}

\newcommand*{\fresh}[1]{\myit{fresh}~{#1}}

\newcommand{\heap}{\code{\sigma}}
\newcommand{\pure}{\code{\pi}}
\newcommand{\wpure}[1]{\code{Pure(#1)}}

\notation{loc}{\ell}

\let\emptyset\empset

\let\implies\Rightarrow

\newcommand{\rulen}[1]{\notationlink{#1}{\ensuremath{{\bf \scriptstyle #1}}}}
\newcommand{\defrulen}[1]{\notationtarget{#1}{\ensuremath{{\bf \scriptstyle #1}}}}

\newcommand{\veepre}{\ensuremath{\wedge_{sp}}}

\notation{substassign}{{:=}}
\newcommand*{\subst}[3]{\coden{[#1\substassign#2]#3}}

\notation{dspec}{{\ensuremath{\mathrm{\Phi}}}}
\newcommand{\effect}{\dspec}

\notation{flowemp}{\code{emp}}
\notation{flow}{\ensuremath{\mathrm{\theta}}}
\newcommand*{\stage}{\ensuremath{E}}
\newcommand*{\DD}{\ensuremath{D}}

\notation{subsume}{\mathrel{\sqsubseteq}}
\newcommand{\stagesubsume}[2]{\ensuremath{#1}\subsume\ensuremath{#2}}
\newcommand{\stagesubs}{\mathrel{\sqsubseteq}}
\newcommand*{\stagereq}[1]{\coden{\req~#1}}
\newcommand*{\stageens}[1]{\coden{\ens[\_]~#1}}
\newcommand*{\stageensres}[2]{\coden{\ens[#1]~#2}}
\newcommand*{\stagepure}[1]{\coden{#1}}
\newcommand*{\stageho}[2]{\coden{\m{#1}(#2)}}
\newcommand*{\stageseq}[2]{\coden{#1\,{\seq}\,#2}}
\newcommand*{\stagedisj}[2]{\coden{#1\,{\vee}\,#2}}

\notation{stageexists}{\exists}
\newcommand*{\stageex}[2]{\coden{\stageexists\,#1\,{\cdot}\,#2}}
\newcommand*{\stageexi}[1]{\coden{{\stageexists}\,{#1}\,{\cdot}\,}}

\newcommand*{\stagepureexi}[1]{\coden{\stageexists\,#1\,{\cdot}\,}}

\newcommand*{\exi}[1]{\coden{\exists\,#1.\,}}
\newcommand*{\fai}[1]{\coden{\forall\,#1.\,}}

\newcommand*{\eref}[1]{\coden{\m{ref}~#1}}
\newcommand*{\ederef}[1]{\coden{!#1}}
\newcommand*{\elet}[3]{\coden{\m{let}~#1{=}#2~\m{in}~#3}}
\newcommand*{\econs}[2]{#1{::}#2}
\newcommand*{\eif}[3]{\coden{\m{if}~#1~\m{then}~#2~\m{else}~#3}}
\newcommand*{\efun}[3]{\codelamsp{(#1)}{#2}{#3}}
\newcommand*{\efunB}[2]{\codelamspB{(#1)}{#2}{}}
\newcommand*{\tlambda}[2]{\efunB{#1}{#2}}
\newcommand*{\ecall}[2]{#1(#2)}
\newcommand*{\eassert}[1]{\coden{\m{assert}~#1}}
\newcommand*{\eassign}[2]{\coden{#1\,{:=}\,#2}}

\newcommand*{\interp}[1]{\llbracket{#1}\rrbracket}

\newcommand*{\seplogicentail}[2]{#1 \entS #2}
\newcommand*{\seplogicentailf}[3]{#1 \entS #2\sep#3}

\notation{specmodels}{\models}
\newcommand*{\specbigstep}[3]{#1 \notationlink{specmodels}{\leadsto} #2 \specmodels #3}
\newcommand*{\specbigstepO}[7]{#1,#2 \notationlink{specmodels}{\leadsto} #5,#6 \specmodels #4}

\notation{slmodels}{\models}
\newcommand*{\seplogicmodels}[2]{#1\slmodels#2}
\notation{slnotmodels}{\not\models}
\newcommand*{\seplogicnotmodels}[2]{#1\slnotmodels#2}

\newcommand*{\andd}{~\text{and}~}
\newcommand*{\suchthat}{~\text{such that}~}

\notation{bigstepeval}{\leadsto}
\newcommand*{\bigstep}[6]{\ensuremath{#1, #2, #3 \bigstepeval #4, #5, #6}}
\newcommand*{\bigstepres}[5]{\ensuremath{#1, #2, #3 \bigstepeval #4, #5}}

\notation{heapstoreupd}{{:=}}
\newcommand*{\storeupdate}[3]{\coden{#1[#2\heapstoreupd#3]}}
\newcommand*{\heapupdate}{\storeupdate}
\notation{storeremovei}{{\not:=}}
\newcommand*{\storeremove}[2]{\coden{#1[#2\storeremovei]}}

\newcommand*{\res}{\m{res}}

\notation{resnormw}{\m{Norm}}
\newcommand*{\resnorm}[1]{\resnormw(#1)}
\notation{reserr}{{\it{Err}}}
\notation{restop}{\top}
\notation{Res}{R}
\notation{Rese}{R_e}

\notation{pureres}{\pure_\res}

\newcommand*{\fstar}{F\ensuremath{\star}}

\notation{inferrable}{^\circ}
\newcommand\darklavendertt[1]{{\small \color{darklavender}{\ttfamily #1}}}
\newcommand\notinferred[1]{\darklavendertt{#1}}
\newcommand\maybeinferred[1]{{\small \color{darkred}{\ttfamily {$\inferrable$}#1}}}
\newcommand\maybeinferredd[1]{{\small \color{darkred}{\ttfamily #1}}}
\newcommand\lavendertt[1]{{\small \color{darklavender}{#1}}}

\notation{roassert}{\coden{@R}}

\notation{disjunion}{\circ}
\notation{subheap}{\subseteq}

\notation{heapdomain}{\m{dom}}
\newcommand*{\heapdom}[1]{\heapdomain(#1)}

\newcommand{\num}[1]{{\ttfamily \color{purple}{{{\#No}}}}}

\usepackage{thmtools}

\usepackage{thm-restate}

\definecolor{darklavender}{rgb}{0.3, 0.16, 0.4}
\usepackage{mathpartir}
\definecolor{babypink}{rgb}{0.96, 0.76, 0.76}
\definecolor{mossgreen}{rgb}{0.68, 0.87, 0.68}

\usepackage[mathscr]{euscript}
 \let\mathscr\relax%
\usepackage[scr]{rsfso}

\usepackage{hyperref}
\hypersetup{hidelinks}

\definecolor{darkmagenta}{rgb}{0.55, 0.0, 0.55}

\definecolor{darkred}{rgb}{0.55, 0.0, 0.0}
\definecolor{forestgreen}{rgb}{0.13, 0.55, 0.13}%
\usepackage{color}

\definecolor{pblue}{rgb}{0.13,0.13,1}
\definecolor{pgreen}{rgb}{0,0.5,0}
\definecolor{pred}{rgb}{0.9,0,0}
\definecolor{pgrey}{rgb}{0.46,0.45,0.48}
\usepackage[capitalise]{cleveref}
\definecolor{airforceblue}{rgb}{0.36, 0.54, 0.66}

\usepackage{listings}
\usepackage{ifthen}
\newboolean{techreport}
\setboolean{techreport}{true}
\newcommand*{\appref}[1]{\ifthenelse{\boolean{techreport}}{\cref{#1}}{\cref*{#1}~\cite{Darius:TR23}}}

\usepackage[misc]{ifsym}

\begin{document}

\title{Staged Specification Logic for Verifying 
Higher-Order Imperative Programs}

\author{Darius Foo(\Letter)\orcidID{0000-0002-3279-5827} \and Yahui Song\orcidID{0000-0002-9760-5895} \and \\ Wei-Ngan Chin\orcidID{0000-0002-9660-5682}}
\authorrunning{Darius Foo, Yahui Song, Wei-Ngan Chin}
\institute{School of Computing, National University of Singapore, Singapore
\email{\{dariusf,yahuis,chinwn\}@comp.nus.edu.sg}}

\maketitle              %
\begin{abstract}

Higher-order functions and imperative states are language features supported by many mainstream languages.
Their combination is expressive and useful, but complicates specification and reasoning, due to the use of yet-to-be-instantiated function parameters.
One inherent limitation of existing specification mechanisms
is its reliance on {\em only 
two stages}~:~an initial stage to denote the precondition at the start of 
the method and a final stage to capture the postcondition. 
Such two-stage specifications force \emph{abstract properties} to be imposed on unknown function parameters, leading to less precise specifications for higher-order methods. 
To overcome this limitation, we introduce a novel 
extension to Hoare logic that supports {\em multiple stages}
for a call-by-value higher-order language
with ML-like local references. 
Multiple stages allow the behavior of unknown function-type parameters to be captured abstractly as uninterpreted relations; and can also model the repetitive behavior of each %
recursion as a separate stage.
In this paper, we define our staged logic with its semantics, prove its soundness and develop a new automated higher-order verifier, called \heifer, for a core ML-like language.

\end{abstract}

\section{Introduction}
\label{sec:intro}
Programs written in modern languages today are rife with higher-order functions~\cite{DBLP:journals/corr/abs-2206-08849,Zampetti2022AnES}, but specifying and verifying them remains challenging, especially if they contain imperative
effects.
Consider the \ml{foldr} function from OCaml.
Here is a %
good specification for it in Iris~\cite{DBLP:journals/jfp/JungKJBBD18}, a state-of-the-art
framework for higher-order concurrent separation logic that is built using Coq proof assistant.

\vspace{-3mm}
\begin{gather*}
  \fai{P,\m{Inv},f,\m{xs},l}
  \biggl\{\,
  \begin{aligned}
    & (\fai{x,a',\m{ys}} \{ P\ x \sep \m{Inv}\ \m{ys}\ a' \}~f(x, a')~\{ r.\ \m{Inv}\ (x{::}\m{ys})\ r \}) \\
    & \sep \m{isList}\ l\ \m{xs} \sep \m{all~P}\ \m{xs} \sep \m{Inv}\ []\ a
  \end{aligned}
  \,\biggr\} \\
  \m{foldr}\ f\ a\ l \\
  \{ r.\ \m{isList}\ l\ \m{xs} \sep \m{Inv}\ \m{xs}\ r \}
\end{gather*}

While this specification is conventional in weakest-precondition calculi like Iris,
one might argue that that this specification is not
the best possible specification for \ml{foldr}, since it requires
two \emph{abstract properties} \coden{\m{Inv}} and
\coden{\m{P}} to summarize the behaviour of \coden{f}.
Moreover, %
the input list \coden{l} is also immutable, through
the same \coden{\m{isList}} predicate in both its pre- and postcondition. (If mutation of list is allowed, a more complex
\coden{\m{Inv}} with an extra mutated list parameter is required.)

These abstract properties must be correspondingly instantiated
for each instance of \coden{f}, but unfortunately some
usage scenarios (to be highlighted later in Sec~\ref{sec:foldr}) of \ml{foldr} cannot be captured by this particular
pre/post specification of Iris, despite how well-designed it was.
Thus, the conventional pre/post approach to specifying higher-order functions currently suffers from possible {\em loss in precision} in its specifications since the presence of these abstract properties implicitly
{\em strengthens} the preconditions for higher-order
imperative methods.

This paper proposes a new logic, \HSSLDef {\em Higher-Order Staged 
Specification Logic} (\HSSL), for specifying and verifying 
higher-order imperative methods. It is designed for 
automated verification via SMT and uses separation logic  
as its core stateful logic, aiming at more precise specifications of heap-based changes. While we have adopted separation logic
to support heap-based mutations, \HSSL~
may also be used with other base logics, %
such as those using dynamic frames~\cite{implicitDFSmansJP12}. We next provide an overview of our methodology by examples before providing formal details and an experimental evaluation of our proposal.

\section{Illustrative Examples}
\label{sec:overview}
\label{sec:examples}
We provide three examples to highlight the key features of
our methodology.

\vspace{-1em}
\subsection{A Simple Example}
\label{sec:helloexample}

\begin{wrapfigure}{r}{0.39\textwidth}
  \vspace{-10mm}
  \begin{lstlisting}[name=hello]
let hello f x y =
  x := !x + 1;
  let r = f y in
  let r2 = !x + r in
  y := r2;
  r2
\end{lstlisting}
\vspace{-1mm}
\caption{A Simple Example}
\label{fig:hello_Example}
\vspace{-4mm}
\end{wrapfigure}

We introduce the specification logic using a 
simple example (\cref{fig:hello_Example}), to
highlight a key challenge we hope to solve, namely
how should we specify the behavior of \ml{hello} without pre-committing to some abstract property on \ml{f}?
To do that, we can model \ml{f} using an {\em uninterpreted relation}.
We use uninterpreted relation rather than a function here in order
to model both over-approximation and possible side-effect.
Since \ml{f} is effectful and may modify arbitrary state, including the references \ml{x} and \ml{y},
a modular specification of \ml{hello} must express the ordering of the call to \ml{f} with respect to the other statements in it so that the caller of \ml{hello} may reason precisely about its effects.
Therefore, a first approximation is the following specification.
We adopt standard separation logic pre/post assertions and extend them with {\em sequential composition} and {\em uninterpreted relations}.
A final parameter (named as \code{res} here) is added to denote the result
of each staged specification's relation (\code{hello} here), a convention we follow henceforth.

{\small 
\begin{tabbing}
  \  \= \quad \= \coden{\req~\heapto{x}{b} \sep \heapto{y}{\_};~\qquad\quad\quad\quad\quad\quad} \= \quad \= \kill
\> \coden{\stageho{hello}{f,x,y,\res} =} \\
\> \> \coden{\stageexi{a}\stagereq{\heapto{x}{a}}\seq}  \> \text{// Stage 1: requiring \ml{x} be pre-allocated} \\
\> \> \coden{\stageens{\heapto{x}{a{+}1}}\seq}  \> \text{// Stage 2: ensuring \ml{x} is updated} \\
\> \> \coden{\stageexi{r}\stageho{f}{y,r}\seq}  \> \text{// Stage 3: unknown higher-order \ml{f} call} \\
\> \> \coden{\stageexi{b}\stagereq{\heapto{x}{b}} {\sep} \heapto{y}{\_};}  \> \text{// Stage 4: requiring %
\ml{x}, \ml{y} be pre\text{-}allocated} \\
\> \> \coden{\stageensres{\res}{\heapto{x}{b}} {\sep} \heapto{y}{\res}{\wedge}\res{=}b{+}r}  \> \text{// Stage 5: %
\ml{y} is updated, and \ml{x} is unchanged} 
\end{tabbing}}

We can summarize the imperative behavior of \ml{hello} before the call to \ml{f}
with a read from \ml{x}, followed by a write to \ml{x}, as captured by 
Stages 1-2. The same applies to the portion after the call to \ml{f} 
(lines 4-6), but here we only consider the scenario when 
\code{x} and \code{y} are disjoint
\footnote{For simplicity, 
the intersection of specifications \coden{\veepre} that arises 
from disjoint preconditions is omitted (with some loss 
in precision) in the main paper, but its core mechanism is 
briefly described in \appref{sec:disjpre}. }. 
Stages 4 and 5 state that memory location \ml{x} is 
being read while \ml{y} will be correspondingly updated.

The ordering of the unknown \ml{f} call with respect to the parts before and after does matter, so the call can be seen as \emph{stratifying} the temporal behavior of the function into \emph{stages}.
Should a specification for \ml{f} become known, usually at a call site,
its \emph{instantiation} may lead 
to a staged formula with only \req/\ens~stages;
which can always be \emph{compacted} into a single \req/\ens\ pair.
We detail a \emph{normalization} procedure to do this in \cref{sec:norm}.

As mentioned before, \ml{f} can modify \ml{x} despite not having direct access to it via an argument, as it could capture \ml{x} from the environment of the caller of \ml{hello}.
To model this, we make worst-case assumptions on the footprints of unknown functions, resulting in the 
precondition $\heapto{x}{b}$ in stage 4. 

\subsection{Pre/Post vs Staged Specifications via \ml{foldr}}
\label{sec:foldr}

We now specify \ml{foldr} and compare it 
with the Iris specification from \cref{sec:intro}.

\begin{center}
\begin{tabular}{c}
\noindent\begin{minipage}{.45\textwidth}
\begin{lstlisting}[name=foldr_sum_state]
let rec foldr f a l =
  match l with
  | [] => a
  | h :: t =>
    f h (foldr f a t)
\end{lstlisting}
\end{minipage}\hfill
\begin{minipage}{.53\textwidth}
{\begin{tabbing}
\quad \= \quad \= \coden{\req~\heapto{x}{b} \sep \heapto{y}{\_};~\quad} \= \quad \= \kill
\coden{\stageho{foldr}{f,a,l,rr} =} \\
\> \> \coden{\stageensres{rr}{l{=}[]{\wedge}rr{=}a}} \\
\> $\vee$ \> \coden{\stageexi{x,r,l_1}} \= \coden{\stageens{l{=}x{::}l_1}\seq} \\
\> \> \> \coden{\stageho{foldr}{f,a,l_1,r}\seq\stageho{f}{x,r,rr}}
\end{tabbing}}
\end{minipage}
\end{tabular}
\end{center}

We model \ml{foldr} as a recursive predicate whose body is a staged formula.
The top-level disjunction represents the two possible paths that result from pattern matching.
In the base case, when \code{l} is the empty list, and the result of \ml{foldr} is \code{a}.
In the recursive case, when \code{l} is nonempty, the specification expresses that the behavior of \ml{foldr} is given by a recursive call to \code{foldr} on the tail of \code{l} to produce a result \code{r}, followed by a call to \code{f} with \code{r} to produce a value for \code{res}.
Crucially, we are able to represent the call to the unknown function \code{f} {directly in the specification}, without being forced
to impose a stronger precondition on \code{f}. 

\code{foldr}'s
specification's
is actually
very precise, %
to the point of mirroring the \ml{foldr} program.
Nevertheless, abstraction may readily be recovered by proving that this predicate entails a weaker formula,
and a convenient point for this would be when the unknown function-typed parameter is instantiated at each of \ml{foldr}'s call sites;
we discuss an example of this shortly.
The point of specifying \ml{foldr} this way is that the precision of stages enables us not to have to commit
to an abstraction prematurely.
We should, of course, summarize
as early as is appropriate
to keep our proving process tractable.

Recursive staged formulae are needed
mainly
to specify higher-order functions with unknown function-typed parameters.
Otherwise, our preference is to apply summarization
to obtain non-recursive staged formulae whenever
unknown function-type parameters have been suitably 
instantiated. Under this scenario, we may still
use recursive pure predicates or recursive
shape predicates in order to obtain best possible 
modular specifications for our program code.

Now, we show how the staged specification for \ml{foldr} 
can be used by
proving that we can sum a list by folding it.
\ml{sum} can be specified in a similar 
way to \ml{foldr}, but since this is a pure function that 
can be additionally checked for termination, we can 
automatically convert it into a {\em pure predicate} 
(without any stages or imperative side effects) to be used 
in (the pure fragment of) our specification logic.
Termination of pure predicates is required for them to be safely used in specifications. (Techniques to check 
for purity and termination are well-known and thus omitted.)
Also, each pure predicate may be used as either a staged predicate
or a pure predicate.
In case a pure predicate \code{p(v^*,\res)} is used as a 
staged predicate; its staged definition is:
\begin{align*}
  \stageho{p}{v^*,\res}  =~ & \stagereq{\emp{\wedge}pre(v^*)}\seq\stageens{\emp{\wedge}\stageho{p}{v^*,\res}}
\end{align*}
\noindent
where \code{pre(v^*)} denotes the precondition 
to guarantee termination and avoids exceptions. 
Note that
$\stageho{p}{v^*,\res}$ is %
overloaded to be used 
as either a staged predicate or a pure predicate. 
This is unambiguous from the context of its use.

\begin{center}
  \vspace{-1mm}
\begin{tabular}{c}
\noindent\begin{minipage}{.45\textwidth}
\begin{lstlisting}[name=foldr_sum_state]
let rec sum li = 
  match li with 
  | [] -> 0
  | x :: xs -> x + sum xs
\end{lstlisting}
\end{minipage}\hfill
\begin{minipage}{.4\textwidth}
{\begin{tabbing}
\quad \= \quad \= \coden{\req~\heapto{x}{b} \sep \heapto{y}{\_};~\quad} \= \quad \= \kill
\coden{\stageho{sum}{li,\res} =} \\
\> \> \coden{\stagepure{l{=}[]{\wedge}\res{=}0}} \\
\> $\vee$ \> \coden{\stagepureexi{r,l_1}\stagepure{l{=}x{::}l_1}{\wedge}\stageho{sum}{l_1,r}{\wedge}\stagepure{\res{=}x{+}r}}
\end{tabbing}}
\end{minipage}
\end{tabular}
\vspace{-1mm}
\end{center}

We can now
re-summarize an imperative use of \code{foldr}
with the help of \code{sum}.

\begin{center}
  \vspace{-1mm}
\begin{tabular}{c}
\begin{minipage}{0.88\linewidth}
\begin{lstlisting}[name=foldr_sum_state]
let foldr_sum_state x xs init
(@\notinferred{$\neffectcall{\m{foldr\_sum\_state(x,xs,init,res)}}= \stageexi{i,r}\stagereq{\heapto{x}{i}}\seq\stageensres{\res}{\heapto{x}{i{+}r}{\wedge}\res{=}r{+}\m{init}{\wedge}\stageho{sum}{xs, r} }$} @)
= let g c t = x := !x + c; c + t in foldr g xs init
\end{lstlisting} 
\end{minipage}
\end{tabular}
\vspace{-1mm}
\end{center}
This summarization gives rise to the following entailment:
\begin{align*}
\quad  &\fai{\m{m},\m{xs},\m{init},\res}\stageho{foldr}{g,xs,init,\res} \\
 \stagesubs \quad &  \stageexi{i,r}\stagereq{\heapto{x}{i}}\seq\stageensres{\res}{\heapto{x}{i{+}r}{\wedge}\res{=}r{+}init {\wedge}\stageho{sum}{xs,r} }
\end{align*}

We have implemented a proof system for subsumption (denoted by $\stagesubs$) between staged formulae %
in our verifier, called \heifer.
This particular entailment can be proved automatically by induction on \ml{xs}. While Iris's earlier pre/post specification for
\code{foldr} can handle this
example through a suitable instantiation of
\coden{(\m{Inv}~\_~\_)}, it
 is \underline{unable} %
to handle the following three %
other call instances.

\begin{center}
  \vspace{-1mm}
\begin{tabular}{c}
\begin{minipage}{0.96\linewidth}
\begin{lstlisting}[name=foldr_sum_state]
let foldr_ex1 l = foldr (fun x r -> let v = !x
                                    in x := v+1; v+r) l 0
let foldr_ex2 l = foldr (fun x r -> assert(x+r>=0);x+r) l 0
let foldr_ex3 l = foldr (fun x r -> if x>=0 then x+r
                                    else raise Exc()) l 0
\end{lstlisting} 
\end{minipage}
\end{tabular}
\vspace{-1mm}
\end{center}

The first example cannot be handled since Iris's current specification for \ml{foldr} expects
its input list \ml{l} to be immutable. The second example fails since
the precondition
required cannot be expressed using just the abstract property
\coden{(P~x)}. The last example fails because the abstract property
\coden{(\m{Inv}~(x::ys)~r)} used in the postcondition of \ml{f} expects
its method calls to return normally.
In contrast, using our approach via staged specification%
, we can re-summarize the
above three call instances to use the following subsumed specifications.

{
  \vspace{-1mm}  
\small\begin{tabbing}
\quad \= \kill
\> \coden{\stageho{foldr\_ex1}{l,res}} \= \coden{\stagesubs} \= \coden{\stageexi{xs}}\code{\stagereq{List(l,xs)}} 
{\seq}
\coden{\stageexi{ys}} \\
\> \> \> \coden{\stageensres{\res}{List(l,ys){\wedge}mapinc(xs,ys){\wedge}sum(xs,res)}} \\
\> \coden{\stageho{foldr\_ex2}{l,res}} \> \coden{\stagesubs} \> \code{\stagereq{allSPos(l)}} 
{\seq}
\coden{\stageensres{\res}{sum(l,res)}} \\
\> \coden{\stageho{foldr\_ex3}{l,res}} \> \coden{\stagesubs} \>
\coden{\stageensres{\res}{allPos(l){\wedge}sum(l,res)}~\vee~(\stageens{{\neg}allPos(l)}\seq Exc())} 
\end{tabbing}
\vspace{-1mm}
}

Note that the first example utilizes a recursive
spatial \coden{\m{List}(l,xs)} predicate, while the last example
used \code{Exc()} as a relation to model
exception as a stage in our specification.
The three pure predicates and one spatial predicate
used in the above can be formally defined, as shown below.

{\small\begin{tabbing}
 \= \kill
\> \coden{\stageho{mapinc}{xs,ys}} \= =~ \coden{(\stagepure{xs{=}[]{\wedge}ys{=}[]})} 
$\vee$ \coden{(\stagepureexi{x,xs_1,ys_1}\stagepure{xs{=}x{::}xs_1{\wedge}ys{=}(x{+}1){::}ys_1}}\\
\> \> \phantom{XXXXXXXXXXXXXXXXXXXX}\code{{\wedge}~\stageho{mapinc}{xs_1,ys_1})} \\
\> \coden{\stageho{allPos}{l}} \> =~ \coden{(\stagepure{l{=}[]})} 
$\vee$ \coden{(\stagepureexi{x,l_1}\stagepure{l{=}x{::}l_1}{\wedge}\stageho{allPos}{l_1}{\wedge}\stagepure{x{\geq}0})} \\
\> \coden{\stageho{allSPos}{l}} \> =~ \coden{(\stagepure{l{=}[]})} 
$\vee$ \coden{(\stagepureexi{x,r,l_1}\stagepure{l{=}x{::}l_1}{\wedge}\stageho{allSPos}{l_1}{\wedge}\stageho{sum}{l,r}{\wedge}\stagepure{r{\geq}0})} \\
\> \coden{\stageho{List}{l,rs}} \> =~ \coden{(\stagepure{\emp{\wedge}l{=}[]})} 
$\vee$ \coden{(\stagepureexi{x,rs_1,l_1} \heapto{x}{r} {\sep} \stageho{List}{l_1,rs_1} {\wedge} \stagepure{l{=}x{::}l_1}{\wedge}\stagepure{rs{=}r{::}rs_1})} 
\end{tabbing}}

We
emphasize that our proposal for staged
logics is strictly more {\em expressive} than
traditional two-stage pre/post specifications, since 
the latter can be viewed as an instance of staged logics.
As an example, the earlier two-stage specification
for \ml{foldr} can be modelled non-recursively in our staged logics as:

\vspace{2mm}
\begin{minipage}{.63\textwidth}
{
\small\begin{tabbing}
\quad \= \quad \= \coden{\req~\heapto{x}{b} \sep \heapto{y}{\_};~\quad} \= \quad \= \kill
\coden{\stageho{foldr}{f,a,l,\res} =} \\
\> \coden{\stageexi{P,Inv,xs}\stagereq{\stageho{List}{l,xs}{\sep}\stageho{Inv}{[],a}{\wedge}\stageho{all}{P,xs}}}  \\
\> \> \coden{\wedge\stageho{f}{x,a',r}{\stagesubs}(\stageexi{ys}\stagereq{\stageho{Inv}{ys,a'}{\wedge}\stageho{P}{x}}{\seq}\,\stageensres{r}{\stageho{Inv}{x{::}ys,r}})\,\seq} \\
\> \coden{\stageensres{\res}{\stageho{List}{l,xs}{\sep}\stageho{Inv}{xs,\res}}}
\end{tabbing}}
\end{minipage}

\subsection{Inferrable vs User-provided Specifications via \code{map}}

\begin{figure}[!h]
\begin{minipage}{.48\textwidth}
\begin{lstlisting}
let rec length xs =
(@ \maybeinferred{$\neffectcall{\m{length(xs,res)}} = ...$} @)
  match xs with
  | [] -> 0
  | x :: xs1 ->
    1 + length xs1

let rec incrg init li =
(@\maybeinferred{$\neffectcall{\m{incrg(init,li,res)}} = ...$}@)
  match li with 
  | [] -> []
  | x :: xs -> init ::
      incrg (init + 1) xs
\end{lstlisting}
\end{minipage}
\begin{minipage}{.5\textwidth}
\begin{lstlisting}[firstnumber=last]
let rec map f xs =
(@\maybeinferred{$\neffectcall{\m{map(f,xs,res)}} = ...$}@)
  match xs with
  | [] -> []
  | x :: xs1 ->
    f x :: map f xs1

let map_incr xs x
(@\notinferred{$\neffectcall{\m{map\_incr}(xs,x,r)}=\\\stageexi{i}\stagereq{\heapto{x}{i}}\seq~\stageexi{m}\stageensres{r}{\heapto{x}{i{+}m}} \\
\phantom{}~{\wedge}~\stageho{length}{xs,m}{\wedge}\stageho{incrg}{i{+}1,\m{xs},r}$}@)
= let f a = x := !x+1; !x
  in map f xs
\end{lstlisting}
\end{minipage}
\vspace{-1mm}
\caption{Implementation of \ml{map\_incr} with a Summarized Specification from \ml{map}} 
\label{fig:map_example_cycle}
\vspace{-4mm}
\end{figure}

Our methodology for higher-order functions is further 
explicated by %
the 
\toelim{ubiquitous} \ml{map} method, shown in \cref{fig:map_example_cycle}. 
Specifications typeset in \notinferred{lavender} must be user-supplied, 
whereas those shown in \inferrableDef \maybeinferred{red} (with the small circle) may be automated or inferred (using the rules of \cref{sec:Hoare}).
Like \ml{sum} before, \ml{length} and \ml{incrg} may be viewed
as \emph{ghost} functions, written only for their specifications to be used to describe behavior. These specifications are also routine and can be mechanically derived; we elide them here and provide them in \appref{sec:elidedspecsexamples}.
The method \ml{map\_incr} describes the scenario we are interested in, where \emph{the state of the closure affects the result of map}.
Its specification states that the pointer \code{x} must have its value incremented by the length of \code{xs}. %
Moreover, 
the \emph{contents} of the resulting list
is captured by another  pure function \ml{incrg},
which builds a list of as many increasing values as there are elements in its input list.

These examples illustrate the methodology involved with 
staged specifications. They inherit the modular verification 
and biabduction-based\cite{DBLP:conf/popl/CalcagnoDOY09}
specification inference of separation 
logic, adding the ability to describe imperative behavior 
using function stages to the mix; biabduction then doubles 
as a means to normalize and compact stages.
\wnnay{clarify and cite bi-abduction}
There is emphasis on the inference of specifications 
and proof automation, and proofs are built out of simple lemmas,
which help summarize behavior and the shapes of data, and either remove 
recursion or move it into a pure ghost function where 
it is easier to comprehend.

In summary, staged logic for specifying imperative higher-order functions
represents a fundamentally new approach that
is %
{\em more general}  %
and yet can be {\em more precise} than what is currently
possible via state-of-the-art
pre/post specification logics for imperative higher-order  methods.
Our main technical contributions to support this new approach include:

\begin{enumerate}[leftmargin=*,labelindent=1pt]
\item \textbf{Higher-Order Staged Specification Logic (\HSSL):}
we design a novel program logic %
to specify the 
behaviors of imperative higher-order methods and give 
its formal semantics.

\item \textbf{Biabduction-based Normalization:}
we propose a normalization procedure for \HSSL~that serves 
two purposes: (i) it allows us to produce succinct staged 
formulae for programs automatically, and (ii) it helps 
structure entailment proof obligations, allowing them to 
be discharged via SMT.

\item \textbf{Entailment:}
we develop a proof system to solve subsumption entailments 
between normalized \HSSL~formulae, prove its soundness, 
and implement an automated prover based on it.

\item \textbf{Evaluation:}
we %
report on initial experimental results, and present various case studies highlighting \HSSL's capabilities. 
 
\end{enumerate}

\section{Language and Specification Logic}
\label{sec:lang}

We target a minimal OCaml-like imperative language with higher-order functions and state.
The syntax %
is given in \cref{fig:core_lang_syntax}. 
Expressions are in ANF (A-normal form); sequencing and control over evaluation order may be achieved using
let-bindings, which define immutable variables.
Mutation may occur through heap-allocated $\m{ref}$s.
Functions are defined by naming lambda expressions, which may be annotated with a specification $\dspec$ (covered below).
For simplicity, they are always in tupled form and their calls are always fully applied.
Pattern matching %
is encoded using recognizer functions (e.g.,~\coden{is\_cons}) and {$\m{if}$} statements.
\coden{\m{assert}}
allows proofs of program properties to be carried out at arbitrary points. \empDef \pointstoDef \sepDef

{ 
  \vspace{-3mm}
\begin{figure}[h]
\centering
 $
  \begin{array}{lrcl}

      \m{({Expressions})} & e & ::= &  
      v
      \vbar x
     \vbar \elet{x}{e_1}{e_2} \vbar
    \ecall{f}{x^*}
    \vbar
    \eref{x} \vbar
    \eassign{x_1}{x_2} \vbar
    \ederef{x}
      \vbar x_1\,{::}\,x_2
    \vbar
   \\
    &&&
    \eassert{\DD}
    \vbar
    \eif{x}{e_1}{e_2}
   \\[0.1em]

      \m{(Values)} &  v & ::= &
      c
      \vbar
      \m{nil}
      \vbar \econs{x_1}{x_2}
      \vbar \efun{x^*}{\dspec[r]}{e}
     \\[0.1em]

      \m{({Staged})} & \code{\dspec}& ::= &
      \coden{\stage}
      \vbar
      \stagedisj{\dspec_1}{\dspec_2}
      \vbar
      \stageseq{\dspec_1}{\dspec_2}
      \vbar
      \stageex{x^*}{\dspec}
     \\[0.1em]

      \m{({Stage})} &  \code{\stage}& ::= &  
      \stagereq{\DD}
      \vbar \stageensres{r}{\DD}
      \vbar \stageho{f}{x^*,r}

      \qquad
      \qquad
      \qquad
      \m{({State})}
      \quad
      \code{\DD}
      ~
      \code{::=}
      ~
      \sigma\wedge\pi

     \\[0.1em]

      \m{({Heap})} &  \code{\sigma}&\code{::=}&  
      \coden{\emp}
      \vbar \coden{\heapto{x_1}{x_2}}
      \vbar \coden{\heap_1\,{\sep}\,\heap_2}

     \\[0.1em]

      \m{({Pure})} &  \code{\pi}&\code{::=}&  
        \coden{\m{true}}
        \vbar \coden{\pure_1{\vee}\pure_2}
        \vbar \coden{\neg\pure}
        \vbar \coden{\exi{x}\,\pure}
        \vbar \coden{t_1{=}t_2}
        \vbar \coden{a_1{<}a_2}
        \vbar \coden{\dspec_1{\stagesubs}\dspec_2}
       \\[0.1em]
      \m{({A{\operatorname{-}}Terms})} &  \code{a}&\code{::=}& 
       \coden{i} \vbar
       \coden{x} \vbar
       \coden{a_1\,{+}\,a_2} \vbar
       \coden{{-}a}
      \\[0.1em]
      \m{({Terms})} &  \code{t}&\code{::=}& 
      \m{nil}
      \vbar \econs{t_1}{t_2} \vbar
       \coden{c} \vbar
       \coden{a} \vbar f \vbar \tlambda{x^*,r}{\dspec} 
\\[0.8em]
  \end{array}$
$\begin{array}{lrcl}
\multicolumn{4}{c}{
\  \code{ \code{c} \in  \mathbb{B} \cup \mathbb{Z} \cup \textbf{unit} }
           \qquad\quad\   \qquad\quad\  
\code{i \in \mathbb{Z}}
           \qquad\quad\   \qquad\quad\  
\code{x, f, r \in \emph{var}}
          
}\\
 \hline
\end{array}$
 \caption{Syntax of the Core Language and Staged Logics} 
 \label{fig:core_lang_syntax}
 \vspace{-3mm}
\end{figure}
}

Program behavior is specified using \emph{staged formulae} \dspecDef $\dspec$, which are disjunctions and/or sequences of \emph{stages} $\stage$.
A stage is an assertion about program state \emph{at a specific point}.
Each stage %
takes one of three forms: a precondition \reqDef \stagereq{\DD}, a postcondition \ensDef \stageensres{r}{\DD} with a named
result \code{r}, or a \emph{function stage} \stageho{f}{v^*, r},
representing
the specification of a (possibly-unknown) function call.
For brevity,
we
use a context notation
\code{\dspec[r]} 
where \code{r} explictly identifies the final
result of specification \code{\dspec}.
Program states \coden{D} are described using separation logic formulae from the \emph{symbolic heap} fragment~\cite{DBLP:conf/popl/CalcagnoDOY09}, without recursive spatial predicates (for simplicity of presentation).
Most values of the core language are as usual also terms of the (pure) logic; a notable exception is the lambda expression $\efun{x^*}{\dspec[r]}{e}$, which occurs in the logic as \logicallambdaDef $\tlambda{x^*,r}{\dspec[r]}$, without its body.
Subsumption assertions between two staged formulae (Sec~\ref{sec:entail}) are denoted by \code{\dspec_1{\stagesubs}\dspec_2}.

\vspace{-1mm}
\subsection{Semantics of Staged Formulae}

\subsubsection{From Triples to Stages}

\newcommand*{\cstate}{\m{st}}

Staged formulae generalize standard Hoare triples.
The standard partial-correctness interpretation of the separation logic Hoare triple
$\HTriple{P(v^*,x^*)}{e}{\exists y^*{\cdot}Q(v^*,x^*,y^*,\res)}$
where \code{v^*} denote valid program variables and \code{x^*} denote
specification variables (e.g., ghost variables) is that
for all states $\cstate$ satisfying $P(v^*,x^*)$, given a reduction $e, \cstate\leadsto^* v, \cstate'$, if $e,\cstate\not\leadsto^*\m{fault}$, then $\cstate'$ satisfies $\exists y^*{\cdot}Q(v^*,x^*,y^*,\res)$.
The staged equivalent
is $\HTriple{\dspec}{e}{\dspec\seq\stageexi{x^*}\stagereq{P(v^*,x^*)}\seq\stageexi{y^*}\stageens{Q(v^*,x^*,y^*,\res)}}$.
Apart from mentioning the \emph{history} $\dspec$, which remains unchanged, its meaning is identical.
Consider, then, $\HTriple{\dspec}{e}{\dspec\seq\stagereq{P_1}\seq\stageens{Q_1}\seq\stagereq{P_2}\seq\stageens{Q_2}}$ -- an intuitive extension of the semantics of triples is that given $e,\cstate\leadsto^*e_1,\cstate_1$, where $\cstate_1$ satisfies $Q_1$, the extended judgment holds if $\cstate_1$ \emph{further} satisfies $P_2$, and reduction from there, $e_1,\cstate_1\leadsto^*e_2,\cstate_2$, results in a state $\cstate_2$ that satisfies $Q_2$.

While heap formulae are satisfied by program states,
staged formulae (like triples), are satisfied by traces which begin and end at particular states.
Uninterpreted function stages further allow
stages to describe the \emph{intermediate states} of programs in specifications -- a useful ability
in the presence of unknown higher-order imperative functions, as we illustrate in \cref{sec:examples} and \appref{sec:casestudies}. 
To formalize all this,
we
give
a semantics for
staged formulae next.

\wnnay{Can we changed the semantics more general form below}
\newcommand{\floToDsp}{\dspec}
\hide{
\begin{verbatim}
 S ::= E | S;S | S\/S
 E ::= ex x* | req D | ens[r] D| f(x*,r)

 S,H,L |= ex x* ==> S1,H,L  iff  \exists v*. S1=S[(x:=v)^*] /\ dom(S)/\{x*)={} 
\end{verbatim}
}

\vspace{-1em}
\subsubsection{Formal Semantics}
\label{sec:spec_semantics}
We first recall the standard \slmodelsDef semantics %
for separation logic formulae
in \cref{fig:semantics_SL}, which provides a useful starting point.

\begin{figure}[!h]
\vspace{-2em}
\begin{equation*}
\begin{aligned}
&\code{\seplogicmodels{S, h}{\sigma \wedge \pure}}
& \code{\m{iff}}\quad\  
& \code{\interp{\pi}_{S} \andd \seplogicmodels{S, h}{\sigma}}
\\[0.05em]
& \code{\seplogicmodels{S, h}{\emp}}
& \code{\m{iff}}\quad\  
& \code{\heapdom{h}=\emptyset} 
\\[0.05em]
&\code{\seplogicmodels{S, h}{\heapto{x}{y}}}
& \code{\m{iff}}\quad\  
& \code{\heapdom{h} = \s{S(x)} \andd h(S(x))} = \interp{y}_{S}
\\[0.05em]
&\code{\seplogicmodels{S, h}{\sigma_1 \sep \sigma_2}}
& \code{\m{iff}}\quad\  
& \code{\exists h_1 h_2. \ 
h_1 {\disjunion} h_2 = h \suchthat
  \seplogicmodels{S, h_1}{\sigma_1} \andd
 \seplogicmodels{S, h_2}{\sigma_2}}
\\
\end{aligned}
\end{equation*}
\vspace{-2mm}
\caption{Semantics of Separation Logic Formulae}
\label{fig:semantics_SL}
\vspace{-2em}
\end{figure}

Let \code{var} be the set of program variables,
\code{val} the set of primitive values, and
$\m{loc} \subset \m{val}$
the set of heap locations; \locDef $\loc$ is a metavariable ranging over locations.
The models are program states, comprising a \emph{store} of variables $S$, a partial mapping from a finite set of variables to values $\m{var} \rightharpoonup \m{val}$, and the heap $h$, a partial mapping from locations to values $\m{loc} \rightharpoonup \m{val}$.
$\interp{\pi}_{S}$ denotes the valuation of
pure formula ${\pi}$ under store $S$.
\heapdomainDef \code{\heapdom{h}} denotes the domain of heap \code{h}. 
\disjunionDef \code{h_1 {\disjunion} h_2 {=} h} denotes
disjoint union of heaps;
if \code{\heapdom{h_1} {\cap} \heapdom{h_2} = \emptyset}, \code{h_1 {\cup} h_2 = h}. 
We write $h_1{\subheap}h_2$ to denote that \subheapDef $h_1$ is a subheap of $h_2$, i.e.,
$\stageexi{h_3} h_1 {\circ} h_3 {=} h_2$.
$\heapupdate{s}{x}{v}$ and $\storeremove{s}{x}$ stand for store/heap \heapstoreupdDef updates and \storeremoveiDef removal of keys.

\begin{figure}[!h]
  \vspace{-3mm}
\begin{equation*}
\begin{aligned}
&\specbigstep{S,h}{S,h_1,\resnorm{\_}}{\stagereq{\heap{\wedge}\pure}}
& \code{\m{iff}}\spaceT
& h_1 {\subheap} h \andd
\seplogicmodels{S, h_1}{\heap{\wedge}\pure}
\\[0.1em]
&\specbigstep{S,h}{S,h,\reserr}{\stagereq{\heap{\wedge}\pure}}
& \code{\m{iff}}\spaceT  
& \forall h_1 \cdot h_1{\subheap}h \implies
\seplogicnotmodels{S, h_1}{\heap{\wedge}\pure}
\\[0.1em]
&\specbigstep{S,h}{S,h,\Res}{\stagereq{(\heap{\wedge}\pure)\roassert}}
& \code{\m{iff}}\spaceT  
&\specbigstep{S,h}{S,h_1,\Res}{\stagereq{(\heap{\wedge}\pure)}}
\\[0.1em]
& \specbigstep{S,h}{S,h {\disjunion} h_1{,}{\resnorm{r}}}{\stageensres{r}{\heap{\wedge}\pure}}\hspace{-0.28cm}
& \code{\m{iff}}\spaceT  
& 
\seplogicmodels{S, h_1}{\heap{\wedge}\pure} \andd
\heapdom{h_1}{\cap}\heapdom{h}{=}\emptyset
\\[0.1em]
& \specbigstep{S,h}{S_1,h_1,\Res}{\stageho{f}{x^*, r}}
& \code{\m{iff}}\spaceT  
& 
S(f) = \efun{y^*}{\dspec[r']}{e},
\\ &&&
\specbigstep{S,h}{S_1,h_1,\Res}{\subst{r'}{r}{\subst{y^*}{x^*}{\effect}}}
\\[0.1em]
& \specbigstep{S,h}{S_1,h_1,\Res}{\stageex{x}{\floToDsp}}
& \code{\m{iff}}\spaceT  
& 
\exists v \,{\cdot}\, %
\specbigstep{\storeupdate{S}{x}{v},h}{S_1,h_1,\Res}{\floToDsp}
\\[0.1em]
& \specbigstep{S,h}{S_2,h_2,\Res}{\floToDsp_1\seq\floToDsp_2}
& \code{\m{iff}}\spaceT  
& 
\specbigstep{S,h}{S_1,h_1,\resnorm{r}}{\floToDsp_1},
\\ &&&
\specbigstep{S_1,h_1}{S_2,h_2,\Res}{\floToDsp_2}
\\[0.1em]
& \specbigstep{S,h}{S_1,h_1,\restop}{\floToDsp_1\seq\floToDsp_2}
& \code{\m{iff}}\spaceT  
& 
\specbigstep{S,h}{S_1,h_1,\restop}{\floToDsp_1}
\\[0.1em]
& \specbigstep{S,h}{S_3,h_3,\resnorm{r_3}}{\stagedisj{\effect_1}{\effect_2}}
& \code{\m{iff}}\spaceT  
& 
\exists h_1{,}h_2{,}r_1{,}r_2\,{\cdot}\,\specbigstep{S,h}{S_1,h_1,\resnorm{r_1}}{\effect_1} 
\\ &&&
\andd \specbigstep{S,h}{S_2,h_2,\resnorm{r_2}}{\effect_2}, \andd
\\ &&& (S_3,h_3,r_3){\in}\s{(S_1,h_1,r_1),(S_2,h_2,r_2)}
\\[0.1em]
& \specbigstep{S,h}{S_1,h_1,\restop}{\stagedisj{\effect_1}{\effect_2}}
& \code{\m{iff}}\spaceT  
& 
\specbigstep{S{,}h}{S_1{,}h_1{,}\restop}{\effect_1} {\,}or{\,}
\specbigstep{S{,}h}{S_1{,}h_1{,}\restop}{\effect_2}
\\[0.1em]
\end{aligned}
\end{equation*}  
\vspace{-1em}
\caption{Semantics of Staged Formulae}
\label{fig:semantics_stages}
\vspace{-2em}
\end{figure}

\wnnay{Change L to set of addresses consumed by req}

We define the semantics of \HSSL\ formulae in \cref{fig:semantics_stages}.
Let $\specbigstep{S,h}{S_1,h_1,\Res}{\dspec}$
denote the \specmodelsDef \emph{models} relation, 
i.e., starting from the program state with 
store \code{S} and heap \code{h},
the formula 
$\effect$ transforms the state into \code{S_1, h_1},
with an intermediate result \ResDef \code{\Res}.
\code{\Res} is either \resnormwDef \code{Norm(r)} for partial
correctness, \reserrDef \code{\reserr} for precondition failure,
or \restopDef \code{\restop} for \emph{possible} precondition failure
in one of its execution paths. 

When $\effect$ is of the form \stagereq{\heap{\wedge}\pure},
the heap $h$ is split into a heaplet $h_1$ satisfying $\heap{\wedge}\pure$,
which is consumed,
and a frame $h_2$, which is left as the new heap.
\emph{Read-only heap assertions}
\roassertDef $(\heap{\wedge}\pure)\roassert$ under $\req$
check but do not change the heap.

When $\effect$ is of the form \stageens{\heap{\wedge}\pure},
$\heap$ describes locations which are to be %
added to the current heap.
The semantics allows some concrete heaplet $h_1$ that
satisfies $\heap{\wedge}\pure$ (containing new or updated locations) be (re-)added to heap $h$.

When $\effect$ is a function stage $\stageho{f}{x^*, r}$,
its semantics depends on the specification of $f$.
A staged existential \stageexistsDef causes the store to be extended with a binding from \code{x} to an existential value \code{v}.
Sequential composition \code{\dspec_1{\seq}\dspec_2}
results in a failure
\code{\restop} if \code{\dspec_1} does,
while disjunction \code{\dspec_1{\vee}\dspec_2} requires both
branches not to fail.

\vspace{-1mm}
\subsection{Compaction}
\label{sec:norm}

Staged formulae subsume separation logic triples, but triples suffice for many verification tasks,
particularly those without calls to unknown functions, and we would like to recover their succinctness in cases where intermediate states are not required.
This motivates a \equivToDef \normsToDef \emph{compaction} or \emph{normalization} procedure for staged formulae, written $\floToDsp ~\normsTo~ \floToDsp$ (\cref{fig:normrules}).
Compaction is also useful for aligning staged formulae, allowing entailment proofs to be carried out stage by stage; we elaborate on this use in \cref{sec:entail}.

\begin{figure}
\vspace{-3em}
\begin{minipage}{.12\textwidth}
 \small \centering
  \begin{align*}
  \coden{\stageseq{\stageens{\false}}{\floToDsp}} & ~\normsTo~ \coden{\stageens{\false}} 
  \quad~ \\
  \coden{\flowemp\seq{\floToDsp}} & ~\normsTo~ \coden{\floToDsp} \\
  \coden{\floToDsp\seq\flowemp} & ~\normsTo~ \coden{\floToDsp} 
  \end{align*}
\end{minipage}%
\begin{minipage}{.35\textwidth}
 \small \centering
  \begin{align*}
  \coden{\stageseq{\stagereq{\DD_1}}{\stagereq{\DD_2}}} & ~\normsTo~ \coden{\stagereq{(D_1{\sep}D_2)}} \\
  \coden{\stageseq{\stageens{D_1}}{\stageensres{r}{D_2}}} & ~\normsTo~ \coden{\stageensres{r}{(D_1{\sep}D_2)}} 
  \end{align*} 
  \incrHRule{
  \coden{\seplogicentail{\DD_A{\sep}\DD_1}}{\coden{\DD_2{\sep}\DD_F}}
  }{
  \coden{\stageseq{\stageensres{r}{\DD_1}}{\stagereq{\DD_2}}} ~\normsTo~ \coden{\stageseq{\stagereq{\DD_A}}{\stageensres{r}{\DD_F}}}
  }{%
  }
\end{minipage}
\caption{Select compaction rules}
\label{fig:normrules}
\vspace{-2em}
\end{figure}

The three rules on the left simplify flows. A false postcondition ($\ens~\heap{\wedge}\false$) models an unreachable or nonterminating program state, so the rest of a flow may be safely ignored.
\flowempDef \coden{\flowemp} in the next two rules 
is \emph{either} \coden{(\req~\emp{\wedge}true)}
or \coden{(\ens~\emp{\wedge}true)}; either may serve as an identity for flows.
The first two rules on the right merge consecutive pre- and postconditions.
Intuitively, they are sound because symbolic heaps separated by sequential composition must be disjoint to be meaningful -- this follows from the use of disjoint union in \cref{fig:semantics_stages}.
The last rule allows a precondition \coden{\req~\DD_2} to be transposed with a preceding postcondition \coden{\ens~\DD_1}.
This is done using biabduction~\cite{DBLP:conf/popl/CalcagnoDOY09}, which computes a pair of antiframe $\DD_A$ and frame $\DD_F$ such that the antiframe is the new precondition required, and frame is what remains after proving the known precondition. The given
rule assumes that $\DD_1$ and $\DD_2$ are disjoint%
\footnote{More exhaustive
aliasing
scenarios are considered
in \appref{sec:disjpre}.
}.
A read-only $\roassert$ heap assertion
under $\req$ would be handled by matching
but not removing from $\DD_F$ (see \cite{DBLP:conf/oopsla/DavidC11}).

\hide{
\begin{verbatim}
 S ::= E | S;S | S\/S
 E ::= ex x* | req D | ens[r] D| f(x*,r)
 ==norm==>
 S ::= T | S\/S
 T ::= E | T;T
 E ::= ex x* | req D | ens[r] D| f(x*,r)
 ==norm==>
 S ::= T | S\/S
 T ::= E | T;T
 E ::= ex x* req D ex y* ens[r] D | f(x*,r)
 ==norm==>
 S ::= T | S\/S
 T ::= E | E;f(x*,r);T 
 E ::= ex x* req D ex y* ens D 

 S,H,L |= ex x* ==> S1,H,L  iff  \exists v*. S1=S[(x:=v)^*] /\ dom(S)/\{x*)={} 
\end{verbatim}
}
Thus staged formulae can always be compacted into the
following form, consisting of a disjunction of \flowDef \emph{flows} $\flow$ (a disjunction-free staged formula)\footnote{Using further normalization rules such as \coden{(\dspec_1{\vee}\dspec_2){\seq\,}\dspec_3~\normsTo~(\dspec_1;\dspec_3){\vee}(\dspec_2{\seq}\dspec_3})},
each consisting of a prefix of function stages (preceded by a description of the intermediate state at that point), followed by a final pre- and postcondition, capturing any behavior remaining after calling unknown
functions.
\begin{align*}
\coden{\dspec} & \coden{::= \flow \vbar \dspec \vee \dspec} \\
\coden{\flow} & \coden{::= (\stageexi{x^*}{\stagereq{\DD}\seq\stageexi{x^*}\stageens{\DD}\seq{\stageho{f}{v^*, r}}}\ ;)^*\ \stageexi{x^*}{\stagereq{\DD}\seq\stageexi{x^*}\stageens{\DD}}}
\end{align*}

An example of compaction is given below (\cref{fig:normexample}, left). We start at the first two stages of the flow and solve a biabduction problem (shown on the right, with solution immediately below) to infer a precondition for the whole flow, or, more operationally, to ``push'' the \textbf{req} to the left.
We will later be able to rely on the new precondition to know that $a=1$ when proving properties of the rest of the flow.
Finally, we
may combine the two \textbf{ens} stages because
sequential composition guarantees disjointness.
Normalization is \emph{sound} in the sense that it transforms staged formulae without changing their meaning.

\begin{figure}
{
\vspace{-2em}
\small
\begin{align*}
& \coden{\ens~\heapto{x}{1}{\sep}\heapto{y}{2}\seq\req~\heapto{x}{a}\seq\ens~\heapto{x}{a{+}1}} & \\
\equivTo~&\coden{\stagereq{a{=}1}\seq\ens~\heapto{y}{2}\seq\ens~\heapto{x}{a{+}1}} & \seplogicentail{\DD_A{\sep}\heapto{x}{1}{\sep}\heapto{y}{2}}{\heapto{x}{a}{\sep}\DD_F} \\
\equivTo~&\coden{\stagereq{a{=}1}\seq\ens~\heapto{y}{2}{\sep}\heapto{x}{a{+}1}} & \DD_A{=}(a{=}1),~ \DD_F{=}(\heapto{y}{2})
\end{align*}
}
\vspace{-2em}
\caption{An example of compaction}
\label{fig:normexample}
\vspace{-2.5em}
\end{figure}

\begin{restatable}[Soundness of Normalization]{theorem}{normsound}
\label{thm:normsound}
Given
  $\dspec_1\normsTo\dspec_2$,
if\ 
  $\specbigstepO{S}{H}{L}{\dspec_1}{S_1}{H_1,\Res_1}{L_1}$,
then
  $\specbigstepO{S}{H}{L}{\dspec_2}{S_1}{H_1,\Res_1}{L_1}$.
\end{restatable}

\begin{proof}
By case analysis on the derivation of $\dspec_1\normsTo\dspec_2$.
See \appref{sec:norm_sound}.
\end{proof}
\vspace{-1.5em}

\section{Forward Rules for Staged Logics}
\label{sec:Hoare}

\begin{figure}[!h]
  
\[
\begin{array}{cc}
  \incrRule{
  \stagesubsume{\dspec_1}{\dspec_3} \quad \HTriple{\dspec_3}{e}{\dspec_4} \quad \stagesubsume{\dspec_4}{\dspec_2}  
  }{\HTriple{\dspec_1}{\coden{e}}{\dspec_2}
  }{Consequence}
\quad\  
  \incrRule{
  \HTriple{\dspec_1}{e}{\dspec_2}   
  }{\HTriple{\dspec;\dspec_1}{\coden{e}}{\dspec;\dspec_2}
    }{Frame}
\end{array}
\]
\[
\begin{array}{cc}
  \incrRule{\fresh{\res}
  }{\HTriple{\dspec}{x}{\stageseq{\dspec}{\coden{\stageensres{\res}{\res{=}x}}}}}{Var}
\ \  
  \incrRule{\fresh{\res}}{\HTriple{\dspec}{v}{\stageseq{\dspec}{\coden{\stageensres{\res}{\res{=}v}}}}}{Val~(c{\vbar}nil{\vbar}x_1{::}x_2)}
\end{array}
\]
  \incrHRule{\fresh{r}}{\HTriple{\dspec}{\eref{x}}{\stageseq{\dspec}{%
        {\stageensres{r}{\heapto{\coden{r}}{x} %
  }}}}}{Ref}

  \incrHRule{\fresh{a,\res}
  }{\HTriple{\dspec}{\coden{\ederef{x}}}{\stageseq{\dspec}{\stageex{a}{\stageseq{\stagereq{\heapto{x}{a}}}{\stageensres{\res}{\heapto{x}{a}{\wedge}\res{=}a
    }}}}}
  }{Deref}

  \incrHRule{%
  }{\HTriple{\dspec}{\coden{x_1{:=}x_2}}{\stageseq{\stageseq{\dspec}{\stagereq{\heapto{x_1}{\_}}}}{\stageensres{\_}{\heapto{x_1}{x_2}%
    }}}
  }{Assign}

  \incrHRule{
  \HTriple{\dspec\seq\stageensres{\_}
    {x}}{e_1}{\dspec_1[r_1]} \quad \HTriple{\dspec\seq\stageensres{\_}
    {{\neg}x}}{e_2}{\dspec_2[r_2]}
  }{\HTriple{\dspec}{\eif{x}{e_1}{e_2}}{\dspec_1 \vee \subst{r_2}{r_1}{\dspec_2}}
  }{If}

  \incrHRule{
  \fresh{x} \quad \HTriple{\dspec}{e_1}{\dspec_1[r]} \quad
  \HTriple{\subst{r}{x}{\dspec_1}}{e_2}{\dspec_2} 
  }{\HTriple{\dspec}{\elet{x}{e_1}{e_2}}{\stageexi{x}
       {\dspec_2}}
  }{Let}

  \incrHRule{
  \fresh{\res} \quad
  \HTriple{\stageensres{\_}{\wpure{\dspec}}}{e}{\dspec_p[r']} \quad \stagesubsume{(\subst{r'}{r}{\dspec_p})}{%
    {\dspec_s}}
  }{\HTriple{\dspec}{\efun{x^*}{\dspec_s[r]}{e}}{\stageseq{\dspec}{\stageensres{\res}{\res{=}\tlambda{x^*,r}{%
            {\dspec_s}}}}}
  }{Lambda}
  \vspace{-0.8cm}

\[
\begin{array}{cc}
  \incrRule{\fresh{\res}
  }{\HTriple{\dspec}{\ecall{f}{x^*}}{\stageseq{\dspec}{\coden{\stageho{f}{x^*,\res}}}}
  }{Call}
  \qquad 
  \incrRule{
  }{\HTriple{\dspec}{\eassert{\DD}}{%
      \stageseq{\dspec}{\stagereq{\DD\roassert}%
      }
    }
  }{Assert}
\end{array}
\]

  \vspace{-1em}
  \caption{Forward Reasoning Hoare Rules with Staged Logics}
  \label{fig:forward_rules}
  \vspace{-2.2em}
\end{figure}

\wnnay{assert rule may lose precision unless we use D@I - @I may complicate paper. undo this change..}
\wnnay{I added a heap {\sep} frame rule which is diff from history frame - UNSOUND}

To verify that a program satisfies a given specification $\dspec_s$, we utilize a set of rules (presented in \cref{fig:forward_rules}) to compute an abstraction or summary of the program $\dspec_p$, then discharge the proof obligation $\stagesubsume{\dspec_p}{\dspec_s}$ (covered in \cref{sec:entail}), in a manner similar to strongest postcondition calculations.

We make use of the following notations. %
$\_$ denotes an anonymous existentially quantified variable.
$\subst{x}{v}{\dspec}$ denotes the \substassignDef substitution of $x$ with $v$ in $\dspec$, giving priority to recently bound variables.
We lift sequencing from flows to disjunctive staged formulae in the natural way:
$\stageseq{\dspec_1}{\dspec_2} \triangleq \bigvee\s{\stageseq{\flow_1}{\flow_2} \mid \flow_1 \in \dspec_2, \flow_2 \in \dspec_2 }$.

The first two rules in \cref{fig:forward_rules} are structural.
The \rulen{Consequence} rule uses \emph{specification subsumption}
(detailed in \cref{sec:entail}) in place of implication --
a form of behavioral subtyping.
The \rulen{Frame} rule has both
a \emph{temporal} interpretation,
which is that the reasoning rules are compositional with respect to the \emph{history} of the current flow,
and a \emph{spatial} interpretation, consistent with the usual one from separation logic, if one uses the normalization rules (\cref{sec:norm}) to move untouched %
p from the final states of $\dspec_1$ and $\dspec_2$ into the frame $\dspec$.

The \rulen{Val} rule illustrates how the results of pure expressions are tracked, in a distinguished $\res$ variable.
The \rulen{Ref} rule results in a new, existentially-quantified location being added to the current state.
The \rulen{Deref} and \rulen{Assign} rules are similar,
both requiring proof that a named location exists with a value, then respectively either returning the value of the location and leaving it unchanged, or changing the location and returning the unit value.
\rulen{Assert} checks its heap state without
modifying it using the $\roassert$ read-only annotation.
\rulen{If} is where disjunctive formula arises,
and \rulen{Let} sequences expressions, renaming the intermediate result of $e_1$ accordingly.
In the \rulen{Let} rule,
program variable in staged formulae
may extend scope of its corresponding variable in
program code if no name clashes. It becomes existentially
quantified when the local variable is out
of scope. %

The \rulen{Lambda} rule handles function definition
annotated
with a given specification $\dspec_s$.
The body of the lambda is summarized into $\dspec_p$ starting from
pure information \code{\wpure{\dspec}} from its program context.
Its behavior must be subsumed by the given specification.
The result is then the lambda expression itself.

The \rulen{Call} rule is completely trivial, yet perhaps the most illuminating as to the design of \HSSL.
A standard modular verifier would utilize this rule to look up the specification associated with $f$, prove its precondition, then assume its postcondition.
In our setting, however, there is the possibility that $f$ is higher-order, unknown, and/or unspecified.
Moreover, there is no need to prove the precondition of $f$ immediately, due to the use of flows for describing program behaviors.
Both of these point to the simple use of a function stage, which stands for a \emph{possibly-unknown} function call.
Utilizing the specification of $f$, if it is provided, is deferred to the unfolding done in the entailment procedure.

We prove soundness of these rules, which is to say that derived specifications faithfully overapproximate the programs they are derived from.
In the following theorem, $\bigstepres{e}{h}{S}{h_1}{S_1}$ is a standard big-step reduction relation whose definition we leave to \appref{sec:bigstep}.
Termination
is also considered
in \appref{sec:term}.
However,
completeness %
is yet to be established.

\begin{restatable}[Soundness of Forward Rules]{thm}{forwardsound}
\label{thm:forwardsound}
Given
  $\HTriple{\emp}{e}{\dspec}$,
then
  $\forall S,h,S_2,h_1 ~{\cdot}~(\specbigstep{S,h}{S_2,h_1,\resnorm{r}}{\dspec})$
  $ \implies \exists S_1 \cdot
   \bigstepres{e}{h}{S}{\resnorm{v},h_1}{S_1}$ and
  $S_1 \subseteq S_2$ and $S_1(r) = v$. 
\end{restatable}

\begin{proof}
  By induction on the derivation of $\bigstepres{e}{h}{S_1}{\Res_1,h_1}{S_1}$.
  See \appref{sec:forw_sound}.
\end{proof}
\vspace{-1.5em}

\section{Staged Entailment Checking and its Soundness}
\label{sec:entail}

In this section, we outline how entailments of the form
$F\vdash\stagesubsume{\dspec_p}{\dspec_s}$ may be automatically checked.
$F$ denotes heap and pure frames that are propagated
by our staged logics entailment rules.
Our entailment %
is always conducted
over the compacted form where non-recursive
staged predicate definitions are unfolded%
,
while unknown predicates are matched exactly.
Lemmas are also used to try re-summarize
each instantiation of recursive staged predicates
to simpler
forms, where feasible.
As staged entailment ensures that all execution traces that satisfy $\dspec_p$ must also \subsumeDef satisfy $\dspec_s$, we rely on
theory of \emph{behavioral subtyping}~\cite{DBLP:journals/toplas/LeavensN15}
to relate them.
Specifically, we check that {\em contravariance holds} for
pre-condition entailment, while %
{\em covariance holds} for
post-condition entailment, as follows:

\incrHRule{
\fresh{y^*} \qquad \seplogicentailf{F_0\sep D_2}{(\exi{x^*} D_1)}{F} \qquad
F\vdash\stagesubsume{\flow_a}{\flow_c}
}{F_0\vdash(\stageexi{x^*}\stagesubsume{\stagereq{\DD_1}\seq\flow_a)}{(\stageexi{y^*}\stagereq{\DD_2}\seq\flow_c)}
}{EntReq}
\incrHRule{
  \fresh{x^*} \qquad  \seplogicentailf{F_0\sep D_1}{(\exi{y^*} D_2)}{F}
  \qquad F\vdash\stagesubsume{\flow_a}{\flow_c}
}{F_0\vdash(\stageexi{x^*}\stagesubsume{\stageensres{r}{\DD_1}\seq\flow_a)}{(\stageexi{y^*}\stageensres{r}{\DD_2}\seq\flow_c)}
}{EntEns}

More details of staged entailment rules are given in \appref{app:entail}.
Note that we use another entailment over
separation logic $\seplogicentailf{D_1}{D_2}{F_r}$ that can propagate
residual frame, \code{F_r}.
Lastly, we outline the soundness of staged entailemt
against the semantics of staged formulae, ensuring that
all derivations are valid.

\begin{restatable}[Soundness of Entailment]{theorem}{entailsound}
\label{thm:entailsound}
Given
  $ %
  \stagesubsume{\dspec_1}{\dspec_2}$ and
  $\specbigstep{S,h}{\resnorm{r_1},S_1,h_1}{\dspec_1}$,
then
  there exists
  $h_2$~such that~$\specbigstep{S,h}{\resnorm{r_1},S_2,h_2}{\dspec_2}$ where $h_2 \subseteq h_1$.
\emph{(Here, $h_1 \subseteq h_2$ denotes that
~$\exi{h_3} h_1 \disjunion h_3 = h_2$.)}
\end{restatable}

\begin{proof}
  By induction on the derivation of $\stagesubsume{\dspec_1}{\dspec_2}$.
  See \appref{sec:entail_sound}.
\end{proof}

\vspace{-0.6cm}
\section{Implementation and Initial Results}
\label{sec:evaluation}

We prototyped our verification methodology
in a tool named \href{https://github.com/hipsleek/heifer}{\text{\heifer}}.
Our tool %
takes input programs written in a subset of OCaml
annotated with user-provided specifications.
It analyzes input programs to produce normalized staged formulae (\cref{sec:norm}, \cref{sec:Hoare}), which it then translates to first-order verification conditions (\cref{sec:entail}) suitable for an off-the-shelf SMT solver.
Here, our prototype targets SMT encodings via Why3~\cite{Fillitre2013Why3W}.
As an optimization, it uses Z3~\cite{DBLP:conf/tacas/MouraB08} directly for queries which do not require Why3's added features.

\newcommand*{\inexpressible}{\xmark} %
\newcommand*{\untried}{-}%

\newcommand*{\heiferratio}{0.37}
\newcommand*{\cameleerratio}{2.49}
\newcommand*{\prustiratio}{0.73}

\begin{table}[h]
   \vspace{-7mm}
   \caption{\label{tab:experiments}
A Comparison with Cameleer and Prusti. 
{\footnotesize (Programs that are natively inexpressible %
are marked with ``\inexpressible''.
Programs that cannot be reproduced from Prusti's artifact \cite{prustiartifact} are marked with ``\untried'' denoting incomparable. 
We use $T$ to denote the total verification time (in seconds) and $T_{P}$ to record 
the time spent on the external provers.)}
}
\vspace{3mm}
\centering
\renewcommand{\arraystretch}{1}
\setlength{\tabcolsep}{5.7pt}
\begin{tabular}{l|cccc|ccc|ccc}
   \Xhline{2\arrayrulewidth}
   \multicolumn{1}{l|}{} & \multicolumn{4}{c|}{Heifer} & \multicolumn{3}{c|}{Cameleer~\cite{DBLP:conf/cav/PereiraR20}} & \multicolumn{3}{c}{Prusti~\cite{DBLP:journals/pacmpl/WolffBMMS21}} \\
   \cline{2-11}
   \multicolumn{1}{l|}{Benchmark} &
   \LoC & \LoS & $T$ & $T_{P}$ &
   \LoC & \LoS & $T$ &
   \LoC & \LoS & $T$ \\
   \Xhline{2\arrayrulewidth}
map & 13 & 11 & 0.66 & 0.58 & 10 & 45 & 1.25 & & \untried & \\
map\_closure & 18 & 7 & 1.06 & 0.77 & & \inexpressible & & & \untried & \\
fold & 23 & 12 & 1.06 & 0.87 & 21 & 48 & 8.08 & & \untried & \\
fold\_closure & 23 & 12 & 1.25 & 0.89 & & \inexpressible & & & \untried & \\
iter & 11 & 4 & 0.40 & 0.32 & & \inexpressible & & & \untried & \\
compose & 3 & 1 & 0.11 & 0.09 & 2 & 6 & 0.05 & & \untried & \\
compose\_closure & 23 & 4 & 0.44 & 0.32 & & \inexpressible & & & \inexpressible & \\
closure~\cite{svendsen2013modular} & 27 & 5 & 0.37 & 0.27 & & \inexpressible & & 13 & 11 & 6.75 \\
closure\_list & 7 & 1 & 0.15 & 0.09 & & \inexpressible & & & \untried & \\
applyN & 6 & 1 & 0.19 & 0.17 & 12 & 13 & 0.37 & & \untried & \\
blameassgn~\cite{Contract:Findler:ICFP02} & 14 & 6 & 0.31 & 0.28 & & \inexpressible & & 13 & 9 & 6.24 \\
counter & 16 & 4 & 0.24 & 0.18 & & \inexpressible & &11 & 7 & 6.37 \\
lambda & 13 & 5 & 0.25 & 0.22 & & \inexpressible & & & \untried & \\
\hline
& 197 & 73 & & & 45 & 112 & & 37 & 27 & \\
\Xhline{2\arrayrulewidth}
\end{tabular}
\vspace{-5mm}
\end{table}

We have verified a suite of %
programs (\cref{tab:experiments}) involving higher-order functions and closures.
As the focus of our work is to explore a new program logic and subsumption-based verification methodology (rather than to verify existing programs), the benchmarks are small in size, and are meant to illustrate the style of specification and give a flavor of the potential for automation.

\cref{tab:experiments} provides an overview of the benchmark suite.
The first two sub-columns show the size of each program (\LoC) and the number of lines of user-provided specifications (\LoS) required.
The next two give the total wall-clock time taken (in seconds) to verify all functions in each program against the provided specifications, and the amount of time spent in external provers.

The next column shows the same programs verified using Cameleer~\cite{DBLP:conf/cav/PereiraR20,Soares2023AFF}, a state-of-the-art deductive verifier.
Cameleer serves as a good baseline for several reasons: it is representative of the dominant paradigm of pre/post specifications and, like Heifer, targets (a subset of) OCaml. It supports higher-order functions in both programs and specifications~\cite{DBLP:journals/corr/abs-2011-14044}.
The most significant differences between Cameleer and Heifer are that Cameleer does not support \emph{effectful} 
higher-order functions and is intended to be used via the 
Why3 IDE in a semi-interactive way (allowing tactic-like 
\emph{proof transformations}, %
used in the above programs).

The last column %
shows %
results for Prusti~\cite{DBLP:journals/pacmpl/WolffBMMS21}.
Despite %
Rust's ownership type system, we compare it against Prusti because of its state-of-the-art support for mutable closures, highlighting differences below.
While we were able to reproduce the claims made in Prusti's OOPSLA 2021 artifact~\cite{prustiartifact}, we were not able to verify many of our own benchmark programs %
due to two technical reasons, namely
lacking support for Rust's \verb|impl Trait| (to return closures) and ML-like cons lists (which caused timeouts and crashes).
Support for closures is also not yet in mainline Prusti~\cite{prusticlosurescurrent}.
Nevertheless, we verified the programs we could use for the artifact, the results of which are shown in \cref{tab:experiments}. 
All experiments were performed on macOS using a 2.3 GHz Quad-Core Intel Core i7 CPU with 16 GB of RAM. Why3 1.7.0 was used, with SMT solvers Z3 4.12.2, CVC4 1.8, and Alt-Ergo 2.5.2.
The Prusti artifact, a Docker image, was run using Moby 25.0.1.

\paragraph{\bf{User annotations required.}} 
Significantly less specification than code is 
required in Heifer, with an average 
\LoS/\LoC~ratio of \heiferratio.
This is helped by two things: the use of function 
stages in specifications, and the use of biabduction-based normalization, which allows the specifications of functions to be mostly automated, requiring only properties and auxiliary lemmas to be provided.
In contrast, Cameleer's ratio is \cameleerratio, due to the need to adequately summarize the behaviors of the function arguments and accompany these summaries with invariants and auxiliary lemmas.
Two examples illustrating this are detailed in \appref{app:cameleer-ex}.
Prusti's ratio is \prustiratio, but a caveat is that in the programs for it, only closure reasoning was used, without lemmas or summarization.

\paragraph{\bf{Expressiveness.}}
\heifer~is able to express many programs that Cameleer cannot, particularly closure-manipulating ones.
This accounts for the \inexpressible~rows in \cref{tab:experiments}.
While some of these can be verified with Prusti, unlike stages, Prusti's call descriptions do not capture ordering~\cite{DBLP:conf/tacas/DenisJ23,prustiartifact}; an explicit limitation as shown by
the \inexpressible~rows in Prusti's column.
Prusti is able to use history invariants and the ownership of the Rust type system, but this difference is more than mitigated in \heifer~with the adoption of an
expressive staged logic with spatial heap state; more %
appropriate for the weaker (but more general)
type system of OCaml.

\vspace{-0.5em}
\section{Related Work}
\label{sec:related}
\vspace{-0.2em}

The use of sequential composition in specifications goes back to classic theories of program refinement, such as
Morgan's refinement calculus~\cite{Morgan1994TheRC}
and
Hoare and He's Unifying Theories~\cite{He1998UnifyingTO},
as well session types~\cite{DBLP:journals/corr/abs-1208-6483} and logics~\cite{DBLP:conf/aplas/CosteaCQC18}.
It has also been used to
structure verification conditions and
give users 
control over
the order in which they 
are given to provers~\cite{DBLP:conf/fm/GherghinaDQC11}, allowing more reliable
proof automation.
We extend both lines of work,
developing the use of sequential composition as a precise specification mechanism for
higher-order imperative functions, and using it to guide entailment proofs of staged formulae.

Higher-order imperative functions
were classically
specified in program logics using \emph{evaluation formulae}~\cite{DBLP:conf/lics/HondaYB05} and \emph{reference-reachability predicates}~\cite{Yoshida:FOSSACS07}.
The advent of separation logic has allowed for simpler specifications using \emph{invariants} and \emph{nested triples} (\cref{sec:intro}).
These techniques are common in
higher-order
separation logics, such as
HTT~\cite{DBLP:journals/jfp/NanevskiMB08},
CFML~\cite{DBLP:conf/icfp/Chargueraud11}, Iris~\cite{DBLP:journals/jfp/JungKJBBD18} and Steel/Pulse~\cite{DBLP:journals/pacmpl/SwamyRFMAM20}, which are
encoded in
proof assistants (e.g. Coq, \fstar~\cite{DBLP:conf/popl/SwamyHKRDFBFSKZ16}) which do not natively support closures or heap reasoning.
While the resulting object logics are highly expressive, they are much more complex (owing to highly nontrivial encodings) and consequently less automated than systems that discharge obligations via SMT.
We push the boundaries in this area by proposing stages as a new, precise specification mechanism which is compatible with automated verification.

The guarantees of an expressive type system can significantly simplify how higher-order state is specified and managed.
Prusti~\cite{DBLP:journals/pacmpl/WolffBMMS21}
exploits this with
\emph{call descriptions} (an alternative to function stages, as pure assertions saying that a call has taken place with a given pre/post)
and \emph{history invariants}, which rely on the ownership of mutable locations that closures have in Rust.
Creusot~\cite{DBLP:conf/tacas/DenisJ23}
uses a \emph{prophetic mutable value semantics}
to achieve a
similar goal with
pre/post specifications of closures.
Our solution is not dependent on an ownership type system,
applying more generally to languages with unrestricted mutation.

Defunctionalization~\cite{reynolds1972definitional} is another promising means of reasoning about higher-order effectful programs~\cite{DBLP:journals/corr/abs-2011-14044},
pioneered by the Why3-based Cameleer~\cite{DBLP:conf/cav/PereiraR20}. This approach
currently
does not
support
closures.

Our approach to automated verification is currently based
on strict evaluation. It would be interesting to see
how staged specifications can be extended to support
verification of lazy programs, as 
had been explored in~\cite{DanaXu:POPL09} and \cite{Liquid:Haskell14}.

\vspace{-0.4cm}
\section{Conclusion}%
\label{sec:conclusion}
\vspace{-0.5em}
We have explored how best to \emph{modularly specify and verify higher-order imperative programs}. 
Our contributions are manifold: we
propose a new staged specification logic, rules for deriving staged formulae from programs and normalizing them using biabduction, and an entailment proof system.
This forms the basis of a new verification methodology, which we validate with our prototype \heifer.

To the best of the authors' knowledge, this work is the first  to
introduce a {\em fundamental} staged specification mechanism for
verifying higher-order imperative programs {\em without} any presumptions;
being
{\em more concise} (without the need for
specifying abstract properties) and {\em more precise}
(without imposing preconditions on function-typed parameters)
when compared to existing solutions.

\vspace{1em}
\noindent {\bf Acknowledgments}
This research is supported by %
the Ministry of Education, Singapore, under the Academic Research Fund Tier 1 (FY2023) (Project Title: Automated Verification for Imperative Higher-Order Programs).
\bibliography{references}
\bibliographystyle{plain}

\newpage
\section*{Appendix}
\appendix

\section{Inferred specifications}
\label{sec:elidedspecsexamples}

Inferred specifications for the examples in the paper.

Pure predicates that can also be proven terminating are shown below. One of them, namely \ml{range}, has a precondition to ensure termination.%
\begin{align*}
  \neffectcall{\m{length}(\m{xs},\res)} =~ &  \stagepure{\m{xs}{=}[]{\wedge}\res{=}0} \\
  & \vee \stagepureexi{r,x,\m{xs}_1}\stagepure{\m{xs}{=}x{::}\m{xs}_1}{\wedge} \stageho{length}{\m{xs}_1,r}\wedge\stagepure{\res{=}r{+}1 } \\
  \neffectcall{\m{incrg}(\m{init},\m{li},\res)} =~ & \stagepure{\m{li}{=}[]{\wedge}\res{=}[]} \\
  & \vee \stagepureexi{r,x,\m{xs_1}}\stagepure{\m{li}{=}x{::}\m{xs_1}}{\wedge} \stageho{incrg}{\m{init}{+}1,\m{xs_1},r}\wedge\stagepure{\res{=}\m{init}{::}r } \\
  \neffectcall{\m{inc}(x,\res)} =~ & \stagepure{\res{=}x{+}1} \\
  \neffectcall{\m{range}(x,n,res)} =~ & \stagereq{\m{n}{\geq}0}\seq \\
  & \stageens{n{=}0{\wedge}\res{=}[]~\vee~ \stagepureexi{r} n{>}0{\wedge}\stageho{range}{x{+}1,n{-}1,r}{\wedge}res{=}x{::}r}
\end{align*}

Note that the
precondition to ensure termination of each
pure predicate would have been baked into the pure predicate's definition itself.
For example, \code{\stageho{range}{x,n,res}  \implies n{\geq}0}.

Below are staged specifications for higher-order methods that have been inferred. Since the unknown function parameter may be imperative, these higher-order functions are currently only expressible in the staged specification format.

\begin{align*}
  \neffectcall{\m{map}(f,\,\m{xs},\,\res)} =~ & \stageens{\m{xs}{=}\m{[]}\wedge\res{=}[]} \\
  & \vee\stageexi{h,x,\m{xs_1}}\stageens{\m{xs}{=}x{::}\m{xs_1}}\seq\stageho{f}{x,h}\seq\stageexi{t}\stageho{\m{map}}{f,\m{xs_1},t}\seq \\
  & \phantom{{}\vee{}}\stageens{\res{=}h{::}t} \\
  \neffectcall{\m{applyN}(f,\m{x},n,\res)} =~ &\stageens{\m{n}{=}0\wedge\res{=}x} \\
  & \vee\stageens{\m{n}{\neq}0}\seq\stageexi{r}\stageho{f}{x,r}\seq\stageho{\m{applyN}}{f,r,n{-}1,\res} \\
  \neffectcall{\m{take}(f,n,\res)} =~ & \stageens{\m{n}{=}0\wedge\res{=}[]} \\
  & \vee\stageens{\m{n}{\neq}0}\seq\stageexi{r,r2}\stageho{f}{r}\seq\stageho{\m{take}}{f,n{-}1,r2}\seq\stageens{\res{=}r{::}r2} \\
\end{align*}

Below is a staged specification for a first-order imperative method that has been inferred.
Two-staged pre/post
specifications are always possible for first-order imperative method that returns normally.

\begin{align*}
  \neffectcall{\m{integers}(res)} =~ & \stageexi{a}\stagereq{\heapto{x}{a}}\seq\stageens{\heapto{x}{res}{\wedge}res{=}a{+}1} 
\end{align*}

\newcommand*{\examplemutclosure}{\ml{mut\_closure}}
\newcommand*{\examplecounter}{\ml{counter}}

\section{Details on Mutable Closures}
\begin{figure}[h]
\begin{minipage}{0.95\textwidth}
\begin{lstlisting}[name=mutable_closure]
let mut_closure () (@\maybeinferred{$\neffectcall{m\_c(res)} = \stageexi{x}\stageens{\heapto{x}{2}}\wedge\res{=}1$}@) =
  let counter =
    let x = ref 0 in
    fun () (@\maybeinferred{$\neffectcall{\lambda(res)}~\stageexi{i}\stagereq{\heapto{x}{i}}\seq\stageens{\heapto{x}{i{+}1}\wedge\res{=}i}$}@) ->
      let r = !x in x := !x + 1; r
  in
  (@\maybeinferred{\s{$\stageexi{x}\stageens{\heapto{x}{0}}{\wedge}\stageho{counter}{res}{=}\stageexi{i}\stagereq{\heapto{x}{i}}\seq\stageens{\heapto{x}{i{+}1}{\wedge}res{=}i}$}}@)
  counter ();
  (@\maybeinferred{\s{$\stageexi{x}\stageens{\heapto{x}{0}}{\wedge}\stageho{counter}{res}{=}\cdots\seq\stageexi{r}\stageho{counter}{r}$}}@)
  (@\maybeinferred{\s{$\stageexi{x,r}\stageens{\heapto{x}{1}}{\wedge}\stageho{counter}{res}{=}\cdots{\wedge}r{=}0$}}@)
  counter ()
  (@\maybeinferred{\s{$\stageexi{x,r,counter}\stageens{\heapto{x}{1}}{\wedge}\stageho{counter}{res}{=}\cdots{\wedge}r{=}0\seq\stageho{counter}{res}$}}@)
  (@\maybeinferred{\s{$\stageexi{x,r,counter}\stageens{\heapto{x}{2}}{\wedge}\stageho{counter}{res}{=}\cdots{\wedge}r{=}0{\wedge}res{=}1$}}@)
  (@\maybeinferred{\s{$\stageexi{x}\stageens{\heapto{x}{2}}{\wedge}res{=}1$}}@)
\end{lstlisting} 
\end{minipage}
\caption{A higher-order function using a mutable closure} 
\label{fig:example_mutable_closure}
\end{figure}

We provide some verification details on
an effectful higher-order function in \cref{fig:example_mutable_closure}.
Here \examplecounter~is a function that captures a heap-allocated location \ml{x}, which is no longer in scope by the point \examplecounter~ is invoked.
The specification $\neffectcall{\lambda(res)}$ (line 4) compositionally describes the behavior of
the lambda expression in \examplecounter: it updates the value of a known location \ml{x} which initially has some value \ml{i}, which it returns.
Reasoning about its enclosing let-expression (line 3), however, we see that \ml{x} is a new location, distinct from any other, resulting in it being existentially quantified.
This quantifier ensures that \ml{x} may only be modified by calling \examplecounter, and surfaces in the specification $\neffectcall{m\_c}$ at line 1, allowing us to precisely describe its result.

For modularity, we might wish to hide the persistence of locations created by \examplemutclosure~(e.g.,~assuming the presence of a garbage collector). We can do this simply by using the specification $\stageens{\res{=}1}$, which is an (intuitionistic)
weakening of $\neffectcall{m\_c}$.

A subtlety about
$\neffectcall{\lambda(res)}$
is that it does not assume anything about the value of \ml{x}, requiring only that \ml{x} exists as a location.
Thus, even if we swapped lines 2 and 3, changing the scope of \ml{x} and allowing the rest of \examplemutclosure to mutate it, $\neffectcall{\lambda(res)}$
would be unchanged.
This is in contrast to other systems which require additional guarantees to specify the lambda modularly; an example is Prusti~\cite{DBLP:journals/pacmpl/WolffBMMS21}, which leverages the ownership guarantees provided by the Rust type system.

\section{Additional Examples}
\label{sec:casestudies}

This section may be read as an extension to \cref{sec:examples}.
Here we highlight a few more interesting and involved example programs.

\subsubsection{Closure-local state}

\begin{figure}
  \begin{subfigure}[b]{.55\linewidth}
{\tiny
\begin{lstlisting}[name=mapexample,numbers=none]
let two_closures ()
(@\maybeinferred{$\stageexi{i,j}\stageens{\heapto{i}{1}{\sep}\heapto{j}{2}}\wedge\res{=}3$}@)
= let f = let x = ref 0 in
   fun () -> x:=!x+1; !x
  (@\maybeinferred{$\s{\stageexi{x}\stageens{\heapto{x}{0}\wedge\stageho{f}{r}{=}\stageexi{a}\\ \stagereq{\heapto{x}{a}}\seq\stageens{\heapto{x}{r}{\wedge}r{=}a{+}1}}}$} @)
  in let g = let x = ref 0 in
        fun () -> x:=!x+2; !x
  (@\maybeinferred{$\s{\stageexi{x,x_2}\stageens{\heapto{x}{0}{\sep}\heapto{x_2}{0}{\wedge}\stageho{f}{r}{=}\cdots\wedge\stageho{g}{r}{=}\stageexi{a_2}\stagereq{\heapto{x_2}{a_2}}\seq \\ \phantom{XXXX}\stageens{\heapto{x_2}{r}{\wedge}r{=}a_2{+}2}}}$} @)
  in f()+g()
  (@\maybeinferred{$\s{\stageexi{x,x_2}\stageens{\heapto{x}{0}{\sep}\heapto{x_2}{0}{\wedge}\stageho{f}{r}{=}\cdots\wedge\stageho{g}{r}{=}{\cdots}}\seq\stageexi{r_1,r_2}\stageho{f}{r_1}\seq\stageho{g}{r_2}\seq
      \\ \phantom{XXXXXXX} \stageens{res{=}r_1{+}r_2} }$} @)
  (@\maybeinferred{$\s{\stageexi{x,x_2,r_1,r_2}\stageens{\heapto{x}{1}{\sep}\heapto{x_2}{0}{\wedge}\stageho{f}{r}{=}\cdots \\ {\wedge}\stageho{g}{r}{=}{\cdots}{\wedge}r_1{=}1}\seq\stageho{g}{r_2}\seq\stageens{res{=}r_1{+}r_2} }$} @)
  (@\maybeinferred{$\s{\stageexi{x,x_2,r_1,r_2}\stageens{\heapto{x}{1}{\sep}\heapto{x_2}{2}{\wedge}\stageho{f}{r}{=}\cdots \\ {\wedge}\stageho{g}{r}{=}{\cdots}{\wedge}r_1{=}1{\wedge}r_2{=}2{\wedge}res{=}r_1{+}r_2} }$} @)
\end{lstlisting}
}
\caption{Multiple closures} 
\label{fig:closure_local_state_example}
  \end{subfigure}
  \begin{subfigure}[b]{.5\linewidth}
\begin{lstlisting}[name=generators,numbers=none]
let rec range x n =
(@\maybeinferred{$ \neffectcall{range(x,n,res)} = \lavendertt{\stagereq{n{\geq}0}}\seq{\cdots}$}@)
  if n = 0 then []
  else x::range (x+1) (n-1)

let rec take f n =
(@\maybeinferred{$\neffectcall{take(f,n,res)} = \cdots
  $}@)
  if n = 0 then []
  else f()::take f (n-1)

let gen_contents x n
(@\notinferred{$\neffectcall{gen\_contents(x,n,res)} = \stageexi{i}\stagereq{\heapto{x}{i}}\\ {\seq}\,
  \stageens{\heapto{x}{i{+}n}}{\wedge}\stageho{\m{range}}{i, n, res} $}@)
= let integers =
   fun () -> let r = !x in
          x := !x + 1; r
  in take integers n
\end{lstlisting} 
\caption{Closures as generators} 
\label{fig:iterator_example}
  \end{subfigure}
  \caption{More uses of stateful closures}
\end{figure}

\cref{fig:closure_local_state_example} shows a simple extension of \cref{fig:example_mutable_closure}.
The internal states of multiple closures may be faithfully modelled using staged formulae.
Again, the existential quantifiers serve both to restrict access to both $x$ locations (renamed to $i$ and $j$) and as a means of enforcing modularity in specifications.
Specification subsumption provides a way to hide them, if the internal state of the closure does not matter in a particular context.

\subsubsection{Closures as generators}

With mutable closures, we can build \emph{generators} -- ``next-element'' functions of type \texttt{unit} $\to$ \texttt{\textquotesingle a} which represent potentially infinite sequences.
In \cref{fig:iterator_example}, we recast \emph{counter} (from \cref{fig:example_mutable_closure}) as the generator \emph{integers}.
We then prove a lemma \emph{gen\_contents} that uses the ghost functions \emph{range} and \emph{take} that says what \emph{integers} contains, \emph{for all} $n$.
With only the given specification, this proof goes through automatically.

\section{Intersection of Staged Specifications}
\label{sec:disjpre}

Intersection specifications are meant to capture
different call scenarios and have an opposite semantics
from disjunctions from conditional expressions.
They currently arise primarily from a more exhaustive consideration
of aliasing scenarios and it can be used to support proof search
over disjoint %
preconditions.

An example of how intersection specifications
may originate
is from the following more complete consideration
from bi-abduction rule:

\centerline{
  \begin{minipage}{.8\textwidth}
  \centering
  \incrHRule{
  \bigwedge~(\coden{\seplogicentail{\DD_A{\sep}\DD_1}}{\coden{\DD_2{\sep}\DD_F}})
  }{
  \coden{\stageseq{\stageens{\DD_1}}{\stagereq{\DD_2}}} ~\equivTo~ \coden{\bigwedge_{sp}~(\stageseq{\stagereq{\DD_A}}{\stageens{\DD_F}})}
  }{Float~Pre}
\end{minipage}%
}
A concrete application is the following
where \code{x} and \code{y} may
be aliases or not.

\centerline{
\begin{minipage}{.7\textwidth}
  \begin{align*}
  \coden{\stageseq{\stageens{\heapto{x}{a}}}{\stagereq{\heapto{y}{b}}}} & ~\equivTo~ \coden{(\stagereq{\heapto{y}{b}}{\seq}\,\stageens{\heapto{x}{a}})\veepre(\stagereq{x{=}y{\wedge}a{=}b})} \\
  \end{align*}
\end{minipage}
}

We use the symbol \code{\veepre} to %
distinguish it from
\code{\vee} that results from conditional constructs. Our normalization
rules would place \code{\veepre} at outermost construct so that they
may be used to support proof search during forward verification.

\vspace{-0.5cm}
\begin{align*}
\coden{\dspec} & \coden{::= \dspec' \vbar \dspec \veepre \dspec} \\
\coden{\dspec'} & \coden{::= \flow \vbar \dspec' \vee \dspec'} 
\end{align*}

An example of its use is the following
more general
staged specification
of \ml{hello} example from Sec~\ref{sec:helloexample}
where \code{\veepre} is used in Stage~4a/4b.

{\small 
\begin{tabbing}
  \  \= \quad \= \coden{\req~\heapto{x}{b} \sep \heapto{y}{\_};~\quad\quad\quad\quad\quad} \= \quad \= \kill
\> \coden{\stageho{hello}{f,x,y,\res} =} \\
\> \> \coden{\stageexi{a}\stagereq{\heapto{x}{a}}\seq}  \> \text{// Stage 1: requiring \ml{x} be pre-allocated} \\
\> \> \coden{\stageens{\heapto{x}{a{+}1}}\seq}  \> \text{// Stage 2: ensuring \ml{x} is updated} \\
\> \> \coden{\stageexi{r}\stageho{f}{y,r}\seq}  \> \text{// Stage 3: unknown higher-order \ml{f} call} \\
\> \> \coden{\stageexi{b}~(~\stagereq{\heapto{x}{b}} {\sep} \heapto{y}{\_};}  \> \text{// Stage 4a: requiring %
\ml{x}, \ml{y} be pre\text{-}allocated} \\
\> \> \coden{\stageens{\heapto{x}{b}} {\sep} \heapto{y}{\res}{\wedge}\res{=}b{+}r}  \> \text{// Stage 5a: %
\ml{y} is updated, and \ml{x} is unchanged} \\
\> \> \coden{\veepre}~\coden{\stagereq{\heapto{x}{b}} {\wedge} x{=}y \seq}  \> \text{// Stage 4b: requiring \ml{x} be pre-allocated with \ml{x=y}} \\
\> \> \phantom{\coden{\vee}}~\coden{\stageens{\heapto{x}{\res}}{\wedge}\res{=}b{+}r~)}  \> \text{// Stage 5b: %
\ml{y} (and hence \ml{x}) is updated} 
\end{tabbing}}

The intersection operator \code{\veepre} 
can be floated outermost so that its new
staged specifications below
can %
support proof search via Hoare-style verification rules.
{\begin{tabbing}
\quad \quad \= \quad \= \coden{\req~\heapto{x}{b} \sep \heapto{y}{\_}\seq\quad\quad\quad\quad\quad} \= \quad \= \kill
\> \coden{\stageho{hello}{f,x,y,\res} =} \\
\> \> \coden{\stageexi{a}\stagereq{\heapto{x}{a}{\wedge}x{\neq}y}\seq\,}\coden{\stageens{\heapto{x}{a{+}1}}\seq} \coden{\stageexi{r}\stageho{f}{y,r}\seq}   \\
\> \> \coden{\stageexi{b}~\stagereq{\heapto{x}{b}} {\sep} \heapto{y}{\_}\seq}  \coden{\stageens{\heapto{x}{b}} {\sep} \heapto{y}{\res}{\wedge}\res{=}b{+}r}   \\
\> \> \coden{\veepre}~\= \coden{\stageexi{a}\stagereq{\heapto{x}{a}{\wedge}x{=}y}\seq\,}\coden{\stageens{\heapto{x}{a{+}1}}\seq} \coden{\stageexi{r}\stageho{f}{y,r}\seq}   \\
\> \> \> \coden{\stageexi{b}~\stagereq{\heapto{x}{b}}\seq}\,\coden{\stageens{\heapto{x}{\res}}{\wedge}\res{=}b{+}r~}   
\end{tabbing}}

\section{An Example to Illustrate the Forward Rules}
\label{app:forward-ex}
As an example to illustrate the forward (Hoare) rules,
consider the program in \cref{fig:forward_rules_example}, annotated with intermediate specifications generated by the rules.
It defines a location $x$ that is captured by a closure and mutates $x$ through the closure, asserting something about the closure's result at the end.

\begin{figure}[!h]

\begin{lstlisting}
let x = ref 0 in
(@\maybeinferred{\s{\toelim{\stageexi{x}}\stageens{\heapto{x}{0}}}}@)
let f = fun () ->
  x := !x + 1;
  (@\maybeinferred{\s{\stageexi{a}\stagereq{\heapto{x}{a}}\seq\stageens{\heapto{x}{a}\wedge\res{=}a}\seq\stageens{\res=a{+}1}\seq\stagereq{\heapto{x}{\_}}\seq\stageens{\heapto{x}{a{+}1}\wedge\res{=}()}}}@)
  (@\maybeinferred{\s{\stageexi{a}\stagereq{\heapto{x}{a}}\seq\stageens{\heapto{x}{a{+}1}\wedge\res{=}()}}}@)
  !x
  (@\maybeinferred{\s{\stageexi{a}\stagereq{\heapto{x}{a}}\seq\stageens{\heapto{x}{a{+}1}}\seq\stageexi{b}\stagereq{\heapto{x}{b}}\seq\stageens{\heapto{x}{b}\wedge\res{=}b}}}@)
  (@\maybeinferred{\s{\stageexi{a}\stagereq{\heapto{x}{a}}\seq\stageens{\heapto{x}{res}\wedge res{=}a{+}1}}}@)
in
(@\maybeinferred{\s{\toelim{\stageexi{x}}\stageens{\heapto{x}{0}{\wedge}f(res){=}\stageexi{a}\stagereq{\heapto{x}{a}}\seq\stageens{\heapto{x}{res}\wedge\res{=}a{+}1}}}}@)
f ();
(@\maybeinferred{\s{\toelim{\stageexi{x}}\stageens{\heapto{x}{0}{\wedge}f(res){=}\cdots}\seq\stageexi{r_1}\stageho{f}{r_1}}}@)
(@\maybeinferred{\s{\toelim{\stageexi{x}}\stageens{\heapto{x}{1}{\wedge}f(res){=}\cdots}}}@)
let r = f() in
(@\maybeinferred{\s{\toelim{\stageexi{x}}\stageens{\heapto{x}{1}{\wedge}f(res){=}\cdots}\seq\toelim{\stageexi{r}}\stageho{f}{r}}}@)
(@\maybeinferred{\s{\toelim{\stageexi{x}}\stageens{\heapto{x}{2}{\wedge}f(res){=}\cdots{\wedge}r{=}2}}}@)
assert (r = 2);
(@\maybeinferred{\s{\toelim{\stageexi{x}}\stageens{\heapto{x}{1}{\wedge}f(res){=}\cdots{\wedge}r{=}2}\seq\stagereq{r{=}2}\toelim{\seq\stageens{r{=}2}}}}@)
(@\maybeinferred{\s{\toelim{\stageexi{x}}\stageens{\heapto{x}{1}{\wedge}f(res){=}\cdots{\wedge}r{=}2}}}@)
r
(@\maybeinferred{\s{\toelim{\stageexi{x}}\stageens{\heapto{x}{2}{\wedge}f(res){=}\cdots{\wedge}r{=}2}\seq\stageens{\res{=}r}}}@)
(@\maybeinferred{\s{\toelim{\stageexi{x}}\stageens{\heapto{x}{2}{\wedge}f(res){=}\cdots{\wedge}r{=}2{\wedge}\res{=}r}}}@)
(@\maybeinferred{\s{\stageexi{x,f,r}\toelim{\stageexi{x}}\stageens{\heapto{x}{2}{\wedge}f(res){=}\cdots{\wedge}r{=}2{\wedge}\res{=}r}}}@)
(@\maybeinferred{\s{\stageexi{x}\toelim{\stageexi{x}}\stageens{\heapto{x}{2}{\wedge}\res{=}2}}}@)
\end{lstlisting}
\vspace{-2mm}
  \caption{Forward rules example}
  \label{fig:forward_rules_example}
  \vspace{-3mm}
\end{figure}

Line 2 is the result of applying the \rulen{Ref} rule, resulting in a new, existentially-quantified location $x$.
Next, we reason about the lambda expression, starting with an empty history.
Inferring a specification for the assignment involves first applying with \rulen{Deref} rule, then \rulen{Val} (taking $+$ as primitive), then \rulen{Assign}, resulting in line 5.
This is then normalized into the specification at line 6.
The \rulen{Deref} rule is applied again, and the result is normalized.

Saving the specification of the lambda expression $\dspec_f$ for later, we continue outside it to line 11, using the \rulen{Call} rule twice, followed by the \rulen{Assert} rule. The result at line 15 is the inferred specification of the entire program and cannot be normalized further.

With this done, the next step might be to prove that this specification is subsumed by another specification that a user might write, e.g.,~$\stageexi{x}\stageens{\heapto{x}{2}\wedge\res{=}2}$.
In the next section, we detail a procedure to do this.

\section{Comparing with User Annotations in Cameleer}
\label{app:cameleer-ex}
Of the programs that can be handled, \heifer~requires much less specification. As an example, \cref{fig:cameleermapfold} 
illustrate
Cameleer specifications for \ml{map} and \ml{foldr}, which may be compared to the programs given in \cref{sec:overview}.

\begin{figure}[!ht]

\begin{lstlisting}[name=cameleer]
let rec map (f:'a -> 'b) (xs:'a list) =
  match xs with
  | [] -> []
  | x :: xs1 -> f x :: map f xs1
(*@ ys = map f xs
      variant xs
      ensures length ys = length xs
      ensures forall i. 0 <= i < length ys ->
                ys[i] = f (xs[i]) *)

(*@ lemma index_shift: forall x:'a, xs:'a list, i:int.
   1 <= i /\ i < length (Cons x xs) ->
     (Cons x xs)[i] = xs[i-1] *)

let rec foldr ((inv : 'b -> 'a seq -> bool) [@ghost])
  (f : 'a -> 'b -> 'b) (xs : 'a list) (acc : 'b)
= match xs with
  | [] -> acc
  | x :: t -> f x (foldr inv f t acc)
(*@ r = foldr inv f xs acc
      requires inv acc []
      requires forall acc x ys.
                 inv acc ys -> inv (f x acc) (cons x ys)
      variant  xs
      ensures  inv r xs *)
\end{lstlisting}
\caption{\ml{foldr} and \ml{map} in Cameleer~\cite{DBLP:conf/cav/PereiraR20}}
\label{fig:cameleermapfold}
\vspace{-4mm}
\end{figure}

Because of the need to summarize the effects of \ml{f}, \ml{map}'s postcondition uses a quantifier (over sequence indices) to talk about the elements of the input and output lists.
This then necessitates lemmas such as \ml{index\_shift} for relating indices to list destructuring, finally requiring more lines of specification per line of code.
The \ml{foldr} specification is similar to the one for Iris (\cref{sec:intro}), but does not use a higher-order triple, instead requiring a ghost argument for an invariant that must be preserved between calls to \ml{f}.
This approach is representative of many verifiers, including Dafny, WhyML, and vanilla F$\star$.
As mentioned before, this parameterization of the specification with a summary of \ml{f} is nontrivial, in that it cannot be mechanically done for every higher-order function, as this pair of examples shows.

\section{Entailment for Staged Specifications}
\label{app:entail}
Our entailment over staged specification is always conducted
over the compacted form where non-recursive
staged predicate definitions are unfolded and compacted,
while unknown predicates are always matched exactly.
Lemmas are also also used to try re-summarize
each instantiation of recursive staged predicates.

\subsection{Staged (Specification) Subsumption}

\begin{figure}[!b]
\vspace{-1mm}
\[
\begin{array}{cc}
\incrRule{
\stagesubsume{\flow_1}{\dspec_c} \qquad \qquad \stagesubsume{\flow_2}{\dspec_c}
}{\stagesubsume{(\flow_1\vee\flow_2)}{\dspec_c}
}{DisjLeft}
\qquad 
\incrRule{
\stagesubsume{\flow_a}{\flow_i} \quad (i{=}1{\vee}i{=}2)
}{\stagesubsume{\flow_a}{(\flow_1\vee\flow_2)}
}{DisjRight}
\end{array}
\]
\vspace{2mm}
\incrHRule{
\fresh{y^*} \qquad \seplogicentailf{F_0\sep D_2}{(\exi{x^*} D_1)}{F} \qquad
F\vdash\stagesubsume{\flow_a}{\flow_c}
}{F_0\vdash(\stageexi{x^*}\stagesubsume{\stagereq{\DD_1}\seq\flow_a)}{(\stageexi{y^*}\stagereq{\DD_2}\seq\flow_c)}
}{EntReq}
\vspace{2mm}
\incrHRule{
  \fresh{x^*} \qquad  \seplogicentailf{F_0\sep D_1}{(\exi{y^*} D_2)}{F}
  \qquad F\vdash\stagesubsume{\flow_a}{\flow_c}
}{F_0\vdash(\stageexi{x^*}\stagesubsume{\stageensres{r}{\DD_1}\seq\flow_a)}{(\stageexi{y^*}\stageensres{r}{\DD_2}\seq\flow_c)}
}{EntEns}
\vspace{2mm}
\incrHRule{
\fresh{x^*,y^*} \quad \pure_p{=}\xpure(\heap{\wedge}\pure) \quad \pure_p \implies (\exists y^* \,{\cdot}\, x_1^*{=}x_2^*{\wedge}r_1{=}r_2) \quad
  \emp\sep\pure_p\vdash\stagesubsume{\flow_a}{\flow_c}
}{\heap{\wedge}\pure\vdash\big(\stageexi{x^*}\stagesubsume{\stageho{f}{x_1^*, r_1}\big)\seq\flow_a}{\big(\stageexi{y^*}\stageho{f}{x_2^*, r_2}\big)\seq\flow_c}
}{EntFunc}
\vspace{2mm}
\[
\begin{array}{cc}
  \incrRule{
  \xpure(\sigma){\wedge}\pi_1\implies\exi{x^*} \pi_2
}{\sigma{\wedge}\pi_1\entS(\exi{x^*} \emp{\wedge}\pi_2)\sep(\sigma{\wedge}\pi_1)
}{SLBase}
&
\incrRule{
  \DD_1\entS(\exi{x^*} \DD_2 \wedge v_1{=}v_2)\sep F
}{\heapto{y}{v_1}\sep\DD_1\entS(\exi{x^*} \heapto{y}{v_2}\sep\DD_2)\sep F
}{SLMatch}
\end{array}
\]

\vspace{2mm}
\incrHRule{
 \fresh{x} \qquad \DD_1\wedge z{=}x \entS(\exi{y^*} \DD_2 \wedge v_1{=}v_2)\sep F
}{\heapto{z}{v_1}\sep\DD_1\entS(\exi{x y^*}\heapto{x}{v_2}\sep\DD_2)\sep F
}{SLMatchEx}
\caption{Staged Subsumption and Selected Entailment Rules}
\label{fig:entailment_rules}
\end{figure}

We assume staged formulae $\dspec$ are normalized, i.e., in the form described at the end of \cref{sec:norm}.
Normalization both \emph{aligns} and \emph{compacts} staged formulae
and is a crucial ingredient to the entailment procedure, which interleaves normalization and rewriting to prove subsumption stage by stage.
We illustrate this proof system formally in \cref{fig:entailment_rules}, and by example at the end of this section.

The rules \rulen{DisjLeft} and \rulen{DisjRight} reduce disjunctive subsumption to subsumption between flows $F\vdash\flow_i\stagesubs\flow_j$, where $F$ is an assumption $\sigma{\wedge}\pi$ which arises from the propagation of separation logic frames; where unspecified, it is $\emp{\wedge}\m{true}$.

The next three rules check subsumption between the individual stages of flows, matching them pairwise before continuing on their tails.
This matching is possible because flows are normalized.
The rules \rulen{EntReq} and \rulen{EntEns} respectively express the contravariance of {\req} preconditions and covariance of {\ens} postconditions.
Subsumption between stages reduces to standard separation logic entailments of the form $\seplogicentailf{\DD_1}{\DD_2}{F}$, where $F$ is an inferred frame,
and all heaps that satisfy $\DD_1$ also satisfy $\DD_2$.
Notably, inferred frames $F$ propagate forward and become assumptions in the separation logic entailments of next stage;
a motivating example is the subsumption $\stagesubsume{\stagereq{\heapto{x}{1}}\seq\stageens{\heapto{x}{2}}}{\stagereq{\heapto{x}{1}\sep F}\seq\stageens{\heapto{x}{2}\sep F}}$ for any $F$.
The propagation of frames from $\req$ to $\ens$ stages has previously been called \emph{enhanced specification subsumption}~\cite[Section 2.4]{DBLP:conf/popl/ChinDNQ08}, and has been historically used for verification of object-oriented programs; we extend this use to $\ens$ and $\req$ in
multi-stage specifications.

The rule \rulen{EntFunc} requires that function constructors $f$ match, and their arguments and return value are provably equal under the pure assumptions in $F_0$.
Something notable about \rulen{EntFunc} is that only the pure portion of the frame propagates further across function stages, as the effects of instantiated function stages may invalidate any assumptions about the heap.
Heap frames may thus be dropped if we are using
intuitionistic separation logic.
If classical separation logic is adopted,
our subsumption procedure will need to
enforce \coden{\heap{=}\emp} at the start of \rulen{EntRule}
and at the end of our specification subsumption procedure.

Separation logic \entSDef entailments are then reduced into first-order implication using 
the so-called ``crunch, crunch'' approach~\cite{DBLP:series/natosec/OHearn12}, via the next three 
rules: \rulen{SLMatch} reduces matching locations 
on both sides into equalities on their contents, 
\rulen{SLMatchEx} does the same for 
existentially-quantified locations (matching 
locations regardless of name to instantiate 
existentials, and possibly requiring backtracking), 
and \rulen{SLBase} serves as the base case, at 
which point the frame is \xpureDef abstracted to 
first-order logic (via the function $\xpure$), 
and the final proof obligation is checked via SMT. 
We use $\xpure$~%
to soundly approximate the spatial information of a heap formula in first-order logic.
We illustrate its behavior by example: $\xpure(\emp) = \m{true}$ and $\xpure(\heapto{x}{1}\sep\heapto{y}{2}) = x{\neq}\nil \wedge y{\neq}\nil \wedge x{\neq} y$.

\vspace{-1mm}
\subsection{Inductive predicates and unfolding}
\label{sec:inductive}

\vspace{-1mm}
The proof system we propose allows %
inductive predicates, which are useful for specifying data structures (as well) as the behaviors of recursive functions -- examples of these were given in \cref{sec:examples} and later in \appref{sec:casestudies}.
Unlike in classic separation logic, where such predicates are used to describe the shapes of data in the heap, here inductive predicates describe \emph{flows} %
via staged formulae.

An inductive predicate definition is of the form $\m{g}(x^*, r) \triangleq \dspec_g$, where $\m{g}$, $x^*$, and $r$ may occur in $\dspec_g$.
Unfolding an inductive definition simply replaces a function stage $\m{g}(y^*, r_g)$ with $\subst{r}{r_g}{\subst{x^*}{y^*}{\dspec_g}}$
This is the meat of the \rulen{Unfold} rule, which then normalizes the result to $\dspec_u$ before continuing.

\incrHRule{
  \stageexi{u^*}\subst{r}{r_g}{\subst{x}{x_g}{\dspec_g}}\seq\flow_a \equivTo\ \dspec_u\\
  g(x, r) \triangleq \dspec_g \qquad
F_0\vdash\dspec_u\stagesubs\big(\stageexi{w^*}\stageho{f}{x_f, r_f}\big)\seq\flow_c
}{F_0\vdash\stagesubsume{\big(\stageexi{u^*}\stageho{g}{x_g, r_g}\big)\seq\flow_a}{\big(\stageexi{w^*}\stageho{f}{x_f, r_f}\big)\seq\flow_c}
}{Unfold}

\vspace{2mm}

The first premise in the rule above may be satisfied by a previous definition of an inductive predicate, or by a local name bound to a logical lambda expression.
For example, we may unfold $y$ in the flow $\stageens{y{=}\tlambda{x*,r}{\dspec}}\seq\stageho{y}{v*}$.
This underscores the first-class nature of lambda terms in the logic, which is necessary to precisely model higher-order behaviors like currying.
Such logical lambda terms are manipulated mostly by the above rule and are encoded in a form that preserves alpha equivalence before being sent to the SMT solver.

One more ingredient is required for reasoning about recursive functions: the use of induction. Inductive predicates provide the means of specification, and annother piece is the automated application of lemmas
which is left in %
\appref{sec:lemmas}.

\section{Proving Lemmas and Rewriting}
\label{sec:lemmas}
Nontrivial proofs in automated program verifiers are often made possible by user-supplied lemmas.
A typical way to provide such lemmas in verifiers for Hoare logic and two-state separation logic is to make use of the Curry-Howard correspondence. By encoding the lemma $P \implies Q$ as a ghost procedure with precondition $P$ and postcondition $Q$, application of the procedure at a specific point corresponds to applying the lemma to the proof state at that point, and the lemma can be separately proved by writing a body for the procedure.

By analogy to this, we allow \emph{subsumption lemmas} in the system of the form $\forall x^*.\ f(y, r)\stagesubs\flow_q$, where $x^*$ may occur free in both $f(y, r)$ and $\flow_q$.
This may be seen as a specific case of the subsumption relation between two disjunctive staged formulae, but with singleton disjuncts on both sides, and only a single function stage on the left.
The restricted form is sufficient to express the induction hypotheses of subsumptions between recursive functions.

Rewriting is very similar to unfolding -- the one difference is that not all arguments of the function stage on the left are parameters, hence the universal quantifier on some.

With both the unfolding and rewriting rules, the proof system is fully defined.
As an example, we consider proving a property of following function.

\begin{center}
\begin{tabular}{c}
\noindent\begin{minipage}{.58\textwidth}
\begin{lstlisting}[name=applyn]
let rec applyN f x n =
  (@\maybeinferred{$\neffectcall{\m{applyN(f,x,n,res)}} = %
  {\cdots}$}@)
  if n = 0 then x
  else let r = f x in applyN f r (n-1)
\end{lstlisting}
\end{minipage}\hfill
\begin{minipage}{.4\textwidth}
\begin{lstlisting}[name=applyn]
let incr x = x + 1
let summary x n
(@ \notinferred{$\neffectcall{\m{summary(x,n,res)}}=\stagereq{n{\geq}0}\seq\stageens{\res{=}x{+}n}$} @)
= applyN incr x n
\end{lstlisting}
\end{minipage}
\end{tabular}
\end{center}

The proof begins with a suitable induction hypothesis. We infer the following one using heuristics. In general, it would have to be provided by the user as a lemma (for re-summarization).

\[
  \fai{x,n,\res}\stageho{applyN}{incr,x,n,\res}
  \stagesubs
  \stagereq{n{\geq}0}~\stageens{\res{=}x{+}n}
\]

The following proof can then be carried out automatically. it illustrates the general approach:
subsumption proofs go stage by stage, attempting to match function stages by interleaving unfolding and normalization.
This ends with either an unknown function stage on both sides or a simple pre- and postcondition, in which case the separation logic proof obligations are discharged, or a pair of differing unknown function stages, in which case the proof fails.

{\tiny %
  \[\inferrule*[Right=Unfold]
  {\inferrule*[Right=Rewrite]
  {\inferrule*[Right=Unfold]
  {\inferrule*[Right=Normalize]
  {\inferrule*[Right=EntNorm]
  {n \geq 0 \implies \m{true} \\ \res=(x+1)+n-1 \implies \res=x+n }
  {\fbox{\stageens{n > 0 \wedge r=x+1 \wedge \res=r+n-1}} \stagesubs \stagereq{n\geq 0}\seq\stageens{\res=x+n}}}
  {\stageens{n {>} 0}\seq\fbox{(\stageens{r{=}x{+}1})}\seq(\stagereq{n{-}1\geq 0}\seq\stageens{\res{=}x{+}n{-}1}) \stagesubs \stagereq{n{\geq} 0}\seq\stageens{\res{=}x{+}n}}}
  {\stageens{n {>} 0}\seq\stageho{inc}{x, r}\seq\fbox{(\stagereq{n{-}1\geq 0}\seq\stageens{\res{=}x{+}n{-}1})} \stagesubs \stagereq{n{\geq} 0}\seq\stageens{\res{=}x{+}n}}}
  {\fbox{(... $\vee$ \stageens{n > 0}\seq\stageho{inc}{x, r}\seq\stageho{\m{applyN}}{inc, n-1, r, \res})} \stagesubs \stagereq{n\geq 0}\seq\stageens{\res=x+n}}}
    {\stageho{\m{applyN}}{inc, n, x, \res} \stagesubs \stagereq{n\geq 0}\seq\stageens{res=x+n}}
\qquad\quad 
\]}

In the above example, with \fbox{boxes} indicating changes from one step to the next, we unfold $\m{applyN}$ and focus on the inductive case.
After rewriting with the induction hypothesis, we unfold $\m{inc}$ and normalize to compact everything into a single \textbf{req}/\textbf{ens} stage, allowing an application of \rulen{EntNorm} to produce first-order proof obligations.

\section{Soundness}
\subsection{Operational semantics}
\label{sec:bigstep}
To facilitate the following soundness proofs, we define a big-step \bigstepevalDef reduction relation with judgments of the form \bigstep{e}{h}{S}{\Rese}{h_1}{S_1}.
Program states consist of a heap $h$ and store $S$, like in \cref{sec:spec_semantics}.
Outcomes \ReseDef $\Rese ::= \resnorm{v} \mid \reserr$
are more constrained compared to $\Res$ -- evaluation results in a value $v$, not a variable $r$, and there is only a $\reserr$ outcome representing a failure, without \emph{possible failure} as there might be in the specification.

\[
\begin{array}{cc}
\incrRule{
}{
  \bigstep{v}{h}{S}{v}{h}{S}
}{Nil, Const, Lambda}
&
\incrRule{
}{
  \bigstep{\econs{x_1}{x_2}}{h}{S}{\econs{S(x_1)}{S(x_2)}}{h}{S}
}{Cons}
\end{array}
\]

\[
\begin{array}{cc}
\incrRule{
}{
  \bigstep{x}{h}{S}{S(x)}{h}{S}
}{Var}
\incrRule{
  \seplogicmodels{h, S}{\sigma{\wedge}\pi}
}{
  \bigstep{(\eassert{\sigma{\wedge}\pi})}{h}{S}{()}{h}{S}
}{Assert}
&
\end{array}
\]

\incrHRule{
  \bigstep{e_1}{h}{S}{v}{h_1}{S_1}\qquad
  \bigstep{e_2}{h_1}{\storeupdate{S_1}{x}{v}}{v_1}{h_2}{S_2}
}{
  \bigstep{(\elet{x}{e_1}{e_2})}{h}{S}{v_1}{h_2}{\storeremove{S_2}{x}}
}{Let}

\[
\begin{array}{cc}
\incrRule{
  S(f) = \efun{x^*}{\Phi}{e}\qquad S(x^*) = v^*\\
  \bigstep{\subst{x^*}{v^*}{e}}{h}{S}{v_2}{h_1}{S_1}
}{
  \bigstep{\ecall{f}{x^*}}{h}{S}{v_2}{h_1}{S_1}
}{App}
&
\incrRule{
  x_1 \in \heapdom{h}
}{
  \bigstep{(\eassign{x_1}{x_2})}{h}{S}{()}{\heapupdate{h_1}{S(x)}{S(v)}}{S}
}{Assign}
\end{array}
\]

\[
\begin{array}{cc}
\incrRule{
}{
  \bigstep{\ederef{x}}{h}{S}{h(S(x))}{h}{S}
}{Deref}
&
\incrRule{
  l \notin \heapdom{h}
}{
  \bigstep{\eref{x}}{h}{S}{\loc}{\heapupdate{h}{\loc}{S(x)}}{S}
}{Ref}
\end{array}
\]

\[
\begin{array}{cc}
\incrRule{
  S(b){=}\m{true}\qquad
  \bigstep{e_1}{h}{s}{v}{h_1}{S_1}
}{
  \bigstep{(\eif{b}{e_1}{e_2})}{h}{S}{v}{h_1}{S_1}
}{If1}
&
\incrRule{
  S(b){=}\false\qquad 
  \bigstep{e_2}{h}{s}{v}{h_1}{S_1}
}{
  \bigstep{(\eif{b}{e_1}{e_2})}{h}{S}{v}{h_1}{S_1}
}{If2}
\end{array}
\]

\subsection{Soundness of Normalization}
\label{sec:norm_sound}

\normsound*

\noindent Intuitively,
normalization of a staged formula preserves its behavior.

\begin{proof}
By case analysis on the derivation of $\dspec_1\normsTo\dspec_2$.
Most cases follow immediately from the semantics of staged formulae (\cref{fig:semantics_stages}).
For example, given $\normsTo{\stagereq{D_1}\seq\stagereq{D_2}}{\stagereq{D_1\sep\stagereq{D_2}}}$, $h = h_0 \disjunion h_1 \disjunion h_2$, where $\seplogicmodels{h_1}{D_1}$ and $\seplogicmodels{h_2}{D_2}$, and $L_1 = h_1 \disjunion h_2$.

The only remaining case is \rulen{Float~Pre} (\cref{sec:norm}).
Here we reason about heaps instead of heap formulae.
Let $h(D_i)$ be a heap such that $\seplogicmodels{h(D_i)}{D_i}$ for $i \in \s{a,f,1,2}$.
First, it must be the case that $D_a$ describes a heap that is in $h$, otherwise it would be impossible to prove the conclusion.
From this we know $h = h(D_a) \disjunion F$ for some framing heap $F$.
Now, from the second premise, we have $h \disjunion h(D_1) = h(D_2) \disjunion h_1$.
Substituting $h$, we have $(h(D_a) \disjunion F) \disjunion h(D_1) = h(D_2) \disjunion h_1$.
From the first premise, assuming soundness of $\vdash$, we have $h(D_a) \disjunion h(D_1) = h(D_2) \disjunion h(D_f)$.
$h_1$ must thus be $h(D_f) \disjunion F$.
Consider the conclusion. It suffices to prove that $h = h(D_a) \disjunion F$ and $h(D_f) \disjunion F = h_1$, which is exactly what we have.

\end{proof}

\subsection{Soundness of Forward Rules}
\label{sec:forw_sound}

The rules are sound if every derivable triple $\HTriple{\emp}{e}{\dspec}$ is valid.

A staged formula $\dspec$ is valid if: given that it specifies a transition from initial configuration $S,h$ to final configuration $S_2,h_1,\resnorm{r}$, execution of $e$ from the same initial configuration results in a \emph{compatible} final configuration $h_1,S_1,\resnorm{v}$.

Compatibility requires that $S_1\subseteq S_2$ and $S_1(r) = v$.
Intuitively, the heap and result have to be equal, while the final store of $\dspec$ is allowed to contain more bindings due to existentials.

\forwardsound*

\begin{proof}
  By induction on the derivation of $\bigstepres{e}{h}{S_1}{\resnorm{v},h_1}{S_1}$.
  \begin{itemize}
    \item The nil, cons, and constant are straightforward: their reduction has no effect on the heap, $\dspec$ is of the form $\stageensres{\res}{\res{=}v}$, $S_1 = S_2$, and $S_1(\res)=v$.
    \item The variable case is similar: $\dspec$ is of the form $\stageensres{\res}{\res{=}x}$, $S_1 = S_2$, and $S_1(\res)=x$.
    \item The lambda case is immediate from the induction hypothesis on $e$, assuming that entailment is sound.
    \item $e$ is of the form $\eref{x}$.
      Reduction results in a store $S_1 = S$ and heap $h_1 = \heapupdate{h}{\loc}{S(x)}$.
      $\dspec$ is of the form
      $\stageensres{\res}{\heapto{\res}{x}}$,
      with final state $S_2 = S$ and heap $h_1$, if we choose $h_1 = \s{\loc\mapsto S(x)}$.
      We also have $S_1(\res)= v = \loc$.
    \item $e$ is of the form $\eassign{x_1}{x_2}$.
      Reduction results in a store $S_1 = S$ and heap $h_1 = \heapupdate{h}{S(x_1)}{S(x_2)}$.
      Given $\dspec$ is of the form $\stagereq{\heapto{x_1}{\_}}\seq\stageensres{\_}{\heapto{x_1}{x_2}}$,
      reduction removes the heap location $x_1$, then re-adds it via with new value $S(x_2)$, effectively modifying the heap at that the assigned location to $h_1$.
    \item $e$ is of the form $\ederef{x}$.
      Reduction has no effect on the heap and store and results in $v=S(x)$.
      $\dspec$ is of the form $\stageexi{a}\stagereq{\heapto{x}{a}}\seq\stageensres{\res}{\heapto{x}{a}{\wedge}\res{=}a}$,
      reduction removes some heap location $\loc$ with value $v$, then immediately readds it, leaving the heap unchanged.
      The store $S_2$ is extended with $\s{a\mapsto v}$. It is a superset of $S_1$ because $h(S(x)) = h(\loc) = v$.

    \item $e$ is of the form $\eassert{D}$.
      Reduction has no effect on the heap, and reduction of $\dspec$ is similar to the case for $\ederef{x}$.
    \item $e$ is of the form $\elet{x}{e_1}{e_2}$.
      Reduction evaluates $e_1$ to a value $v$, evaluates $e_2$ with $x$ bound to $v$ in the store, then removes the binding for $x$, resulting in a heap $h$ and store $S_1$.
      $\dspec$ is $\dspec=\stageexi{x}\subst{\res}{x}{\dspec_1}\seq\subst{x}{x}{\dspec_2}$, where
      $\HTriple{\emp}{e_1}{\dspec_1}$ and $\HTriple{\emp}{e_2}{\dspec_2}$.
      Given the induction hypotheses, it remains to show that the heap resulting from reducing $\dspec$ is $h$ and the resulting store $S_2$ is a superset of $S_1$.
      The former follows directly from the semantics of staged formulae as $\m{let}$ expressions do not further change the heap.

    \item $e$ is of the form $\eif{x}{e_1}{e_2}$.
      Reduction proceeds with either $e_1$ or $e_2$, depending on the value of $S(x)$. The conclusion follows from case analysis on $x$, then application of either induction hypothesis.
    \item $e$ is of the form $\ecall{f}{x^*}$.
      Its a summary is a simple function stage $\stageho{f}{x^*, res}$.
      From the big-step reduction relation, we know that $e$ evaluates as $\subst{x^*}{S(x^*)}{e_1}$ would,
      given $S(f) = \efun{x^*}{\dspec}{e_1}$.
      Also, from the semantics of staged formulae, it suffices to prove soundness for $\subst{y^*}{x^*}{\effect}$.
      We know that $e_1$ is summarized soundly from the induction hypothesis, and the substitutions do not compromise this as both $e_1$ and $\dspec$ contain (the values of) $x^*$ after.

  \end{itemize}
\end{proof}

\subsection{Soundness of Entailment}
\label{sec:entail_sound}

As we use a standard fragment of separation logic with the usual semantics, we assume the soundness of its entailment proof system.

\begin{theorem}[Soundness of SL entailment]
\label{thm:slsound}
Given
  $\seplogicentail{D_1}{D_2}$ and
  $\seplogicmodels{S,h}{D_1}$
then
  $\seplogicmodels{S,h}{D_2}$.
\end{theorem}

We focus on the soundness of the entailment proof system for staged formulae.
Given a derivation $\stagesubsume{\dspec_1}{\dspec_2}$ and the same starting configuration $S,H$ with an empty local heap, ``executing'' $\dspec_1$ should result in an an ending configuration with a smaller global heap and larger local heap than executing $\dspec_1$.
The relation between heaps is containment rather than equality intuitively because a stronger formula is allowed to entail a weaker one, e.g.,~$\stageens{\heapto{x}{1}\sep\heapto{y}{1}}$ should entail $\stageens{\heapto{x}{1}}$.

\entailsound*

\begin{proof}
  By induction on the derivation of $\stagesubsume{\dspec_1}{\dspec_2}$.
  \begin{itemize}
    \item In \rulen{DisjLeft}, when the derivation is of the form $\stagesubsume{(\floToDsp_1\vee\floToDsp_2)}{\dspec_2}$, we know from the induction hypothesis that execution of both $\floToDsp_1$ and $\floToDsp_2$ result in global heaps $H_3$ and $H_4$ that are larger than $H_2$, and local heaps $L_3$ and $L_4$ that are smaller than $L_2$.
    Since both ending configurations satisfy the property, and the semantics of disjunction allows us to choose either ending configuration, the conclusion is true.

    \item In \rulen{DisjRight}, when the derivation is of the form $\stagesubsume{\floToDsp_a}{(\floToDsp_1\vee\floToDsp_2)}$, we know from the induction hypothesis that only the execution of $\floToDsp_1$ produces an ending configuration that satisfies the property.
    Since the semantics of disjunction allow execution of $\floToDsp_1\vee\floToDsp_2$ to choose either $\floToDsp_1$ or $\floToDsp_2$, it suffices to prove the property for just one disjunct, which we have by the induction hypothesis.

    \item In \rulen{EntReq},
    when the derivation is of the form $F_0 \vdash \stagesubsume{(\stageexi{x^*}\stagereq{D_1}\seq\floToDsp_a)}{(\stageexi{y^*}\stagereq{D_2}\seq\floToDsp_c)}$,
    we must show that $\stagereq{D_2}$ results in an identical store, smaller global heap, and larger local heap than $\stagereq{D_1}$ (and the induction hypothesis ensures that these relations are preserved).
    In other words, $D_2$ should move \emph{more} of the global heap than $D_1$ does into $L_2$ (resp. $L_1$).
    Since we know $\seplogicentail{\emp\sep D_2}{D1}$, from \cref{thm:slsound}, we know that any heap satisfying $D_2$ must satisfy $D_1$.
    $D_2$ must thus be a more precise heap formula, with more conjuncts.
    Thus, it may removed more of the heap than $D_1$, making $H_2$ smaller.
    Correspondingly, $L_2$ will be larger than $L_1$ as a larger portion of the heap was moved into it.
    \textbf{req} does not update the store, hence the conclusion is true.

    \item The \rulen{EntEns} case is dual to \rulen{EntReq}, but largely similar. Given a derivation of the form
    $F_0 \vdash \stagesubsume{(\stageexi{x^*}\stageens{D_1}\seq\floToDsp_a)}{(\stageexi{y^*}\stageens{D_2}\seq\floToDsp_c)}$, we know $\seplogicentail{D_1}{D_2}$. Hence $D_1$ must add a larger portion of heap into the global heap $H_1$ than $D_2$ does, resulting in $H_1$ being larger and $L_1$ being smaller.
    \textbf{ens} does update $\res$ in the store, but the compatibility of the values that the store is updated with is ensured by the validity of the separation logic entailment.

    \item The \rulen{EntFunc} case seeks to prove that both sides of the entailment are effectively equal under the assumptions in $F_0$. As the store and both heaps are modified the same way (by the formula that is the definition of $f$), they remain equal, hence the conclusion is true.

    \item Both rules for unfolding are technically involved, but conceptually simple. As we replace function stages with their specifications (which have been earlier checked for soundness), the conclusion follows from the soundness of normalization (\cref{thm:normsound}).

  \end{itemize}
\end{proof}

\subsection{Termination}
\label{sec:term}
We outline termination arguments for our normalization procedure,
Hoare-style forward rules and also the
entailment/subsumption procedure for staged specifications

\subsection{Termination of Normalization}

The normalization procedure \code{\dspec_1 \normsTo \dspec_2}
is terminating since staged logic expression is
always getting smaller, except for the \coden{Float Pre}
rule, where each pre-condition is being floated outwards.
Hence, by giving lower weights to pre-condition stages, this rule
will continues to decrease the termination
measure of our normalization procedure.Hence, each application of
normalization/compaction will always terminate.

\subsection{Termination of Hoare-Style Forward Rules}

Except for the consequence rule, all the forward
reasoning rules are either just terminating (CVar, Val, Assign,
Call and Assert).
or works on structurally smaller expressions (e.g. Lambda, Let and If Hoare-rules). If we assume that Consequence
rule is applied at most once for each distinct sub-expression,
the set of Hoare-style forward-reasoning rules is always terminating.

\subsection{Termination of Staged Entailment}

Staged Entailment is used mostly for the summarization
of recursive staged predicates. This process requires
users to provide suitable lemmas that must be appropriately
generalized. In case this is not done properly, there
is a possibility that staged entailment may go into
an infinite loop. To prevent this problem, we only
allow a bounded set of new staged predicate definitions
to be defined, and a bound number of unfoldings.
Once these two bounds are exceeeded, we
approximate our staged entailment outcome with
a failure. This approximation is sound but
may lead to incompleteness for our
entailment procedure over staged logics.

\end{document}